\documentclass[preprint,aps,prd,tightenlines,superscriptaddress,nofootinbib]{revtex4}

  \usepackage{amssymb}
   \usepackage{amsmath}
    \usepackage{amsfonts}
\usepackage[dvips,final]{graphicx}
\usepackage{epsfig}
\usepackage{graphicx}
\usepackage{dcolumn}
\date{}

\date{}
\begin{document}

\newcommand{\ds}{\displaystyle}
\newcommand{\mc}{\multicolumn} 
\newcommand{\bce}{\begin{center}}
\newcommand{\ece}{\end{center}}
\newcommand{\beq}{\begin{equation}}
\newcommand{\eeq}{\end{equation}}
\newcommand{\bea}{\begin{eqnarray}}

\newcommand{\eea}{\end{eqnarray}}
\newcommand{\cont}{\nonumber\eea\bea}
\newcommand{\cl}[1]{\begin{center} {#1} \end{center}}
\newcommand{\ba}{\begin{array}}
\newcommand{\ea}{\end{array}}

\newcommand{\ab}{{\alpha\beta}}
\newcommand{\cd}{{\gamma\delta}}
\newcommand{\dc}{{\delta\gamma}}
\newcommand{\ac}{{\alpha\gamma}}
\newcommand{\bd}{{\beta\delta}}
\newcommand{\abc}{{\alpha\beta\gamma}}
\newcommand{\eps}{{\epsilon}}
\newcommand{\lam}{{\lambda}}
\newcommand{\mn}{{\mu\nu}}
\newcommand{\mpnp}{{\mu'\nu'}}
\newcommand{\Amuu}{{A_{\mu}}}
\newcommand{\Amuo}{{A^{\mu}}}
\newcommand{\Vmuu}{{V_{\mu}}}
\newcommand{\Vmuo}{{V^{\mu}}}
\newcommand{\Anuu}{{A_{\nu}}}
\newcommand{\Anuo}{{A^{\nu}}}
\newcommand{\Vnuu}{{V_{\nu}}}
\newcommand{\Vnuo}{{V^{\nu}}}
\newcommand{\Fmnu}{{F_{\mu\nu}}}
\newcommand{\Fmno}{{F^{\mu\nu}}}

\newcommand{\abcd}{{\alpha\beta\gamma\delta}}


\newcommand{\bsigma}{\mbox{\boldmath $\sigma$}}
\newcommand{\btau}{\mbox{\boldmath $\tau$}}
\newcommand{\brho}{\mbox{\boldmath $\rho$}}
\newcommand{\bpipi}{\mbox{\boldmath $\pi\pi$}} 
\newcommand{\bss}{\bsigma\!\cdot\!\bsigma}
\newcommand{\btt}{\btau\!\cdot\!\btau}
\newcommand{\bnabla}{\mbox{\boldmath $\nabla$}}
\newcommand{\bphi}{\mbox{\boldmath $\tau$}}
\newcommand{\bvarphi}{\mbox{\boldmath $\rho$}}
\newcommand{\bDelta}{\mbox{\boldmath $\Delta$}}
\newcommand{\bpsi}{\mbox{\boldmath $\psi$}}
\newcommand{\bPsi}{\mbox{\boldmath $\Psi$}}
\newcommand{\bPhi}{\mbox{\boldmath $\Phi$}}
\newcommand{\bnab}{\mbox{\boldmath $\nabla$}}
\newcommand{\bpi}{\mbox{\boldmath $\pi$}}
\newcommand{\btheta}{\mbox{\boldmath $\theta$}}
\newcommand{\bkappa}{\mbox{\boldmath $\kappa$}}
\newcommand{\bp}{\mbox{\boldmath $p$}}
\newcommand{\bq}{\mbox{\boldmath $q$}}
\newcommand{\br}{\mbox{\boldmath $r$}}
\newcommand{\bs}{\mbox{\boldmath $s$}}
\newcommand{\bk}{\mbox{\boldmath $k$}}
\newcommand{\bl}{\mbox{\boldmath $l$}}
\newcommand{\bb}{\mbox{\boldmath $b$}}
\newcommand{\be}{\mbox{\boldmath $e$}}
\newcommand{\bP}{\mbox{\boldmath $P$}}
\newcommand{\bB}{\mbox{\boldmath $B$}}


\newcommand{\bT}{{\bf T}}
\newcommand{\fph}{${\cal F}$}
\newcommand{\aph}{${\cal A}$}
\newcommand{\dph}{${\cal D}$}
\newcommand{\fpi}{f_\pi}
\newcommand{\mpi}{m_\pi}
\newcommand{\Tr}{{\mbox{\rm Tr}}}
\def\Qb{\overline{Q}}
\newcommand{\delu}{\partial_{\mu}}
\newcommand{\delo}{\partial^{\mu}}
%
%
\newcommand{\up}{\!\uparrow}
\newcommand{\upup}{\uparrow\uparrow}
\newcommand{\updo}{\uparrow\downarrow}
\newcommand{\uu}{$\uparrow\uparrow$}
\newcommand{\ud}{$\uparrow\downarrow$}
\newcommand{\auu}{$a^{\uparrow\uparrow}$}
\newcommand{\aud}{$a^{\uparrow\downarrow}$}
\newcommand{\pu}{p\!\uparrow}

\newcommand{\qp}{quasiparticle}
\newcommand{\sa}{scattering amplitude}
\newcommand{\ph}{particle-hole}
\newcommand{\qcd}{{\it QCD}}
\newcommand{\integ}{\int\!d}
\newcommand{\ie}{{\sl i.e.~}}
\newcommand{\etal}{{\sl et al.~}}
\newcommand{\etc}{{\sl etc.~}}
\newcommand{\rhs}{{\sl rhs~}}
\newcommand{\lhs}{{\sl lhs~}}
\newcommand{\eg}{{\sl e.g.~}}
\newcommand{\ef}{\epsilon_F}
\newcommand{\sigt}{d^2\sigma/d\Omega dE}
\newcommand{\sige}{{d^2\sigma\over d\Omega dE}}
\newcommand{\rpaeq}{\beq
\left ( \begin{array}{cc}
A&B\\
-B^*&-A^*\end{array}\right )
\left ( \begin{array}{c}
X^{(\kappa})\\Y^{(\kappa)}\end{array}\right )=E_\kappa
\left ( \begin{array}{c}
X^{(\kappa})\\Y^{(\kappa)}\end{array}\right )
\eeq}
\newcommand{\ket}[1]{| {#1} \rangle}
\newcommand{\bra}[1]{\langle {#1} |}
\newcommand{\ave}[1]{\langle {#1} \rangle}
\newcommand{\half}{{1\over 2}}

\newcommand{\singlespace}{
    \renewcommand{\baselinestretch}{1}\large\normalsize}
\newcommand{\doublespace}{
    \renewcommand{\baselinestretch}{1.6}\large\normalsize}
\newcommand{\bftau}{\mbox{\boldmath $\tau$}}
\newcommand{\bfalpha}{\mbox{\boldmath $\alpha$}}
\newcommand{\bfgamma}{\mbox{\boldmath $\gamma$}}
\newcommand{\bfxi}{\mbox{\boldmath $\xi$}}
\newcommand{\bfbeta}{\mbox{\boldmath $\beta$}}
\newcommand{\bfeta}{\mbox{\boldmath $\eta$}}
\newcommand{\bfpi}{\mbox{\boldmath $\pi$}}
\newcommand{\bfphi}{\mbox{\boldmath $\phi$}}
\newcommand{\bfR}{\mbox{\boldmath ${\cal R}$}}
\newcommand{\bfL}{\mbox{\boldmath ${\cal L}$}}
\newcommand{\bfM}{\mbox{\boldmath ${\cal M}$}}
\def\dblint{\mathop{\rlap{\hbox{$\displaystyle\!\int\!\!\!\!\!\int$}}
    \hbox{$\bigcirc$}}}
\def\ut#1{$\underline{\smash{\vphantom{y}\hbox{#1}}}$}

\def\UNITY{{\bf 1\! |}}
\def\Pom{{\bf I\!P}}
\def\lsim{\mathrel{\rlap{\lower4pt\hbox{\hskip1pt$\sim$}}
    \raise1pt\hbox{$<$}}}         
\def\gsim{\mathrel{\rlap{\lower4pt\hbox{\hskip1pt$\sim$}}
    \raise1pt\hbox{$>$}}}         
\def\beq{\begin{equation}}
\def\eeq{\end{equation}}
\def\bea{\begin{eqnarray}}
\def\eea{\end{eqnarray}}

\title{Diffractive dissociation of gluons into heavy quark-antiquark
  pairs in proton-proton collisions
}

\author{Marta~\L uszczak}%
\email{luszczak@univ.rzeszow.pl}
\affiliation{Institute of Physics, University of Rzesz\'ow, PL-35-959 Rzesz\'ow, Poland}

\author{Wolfgang~Sch\"afer}%
\email{Wolfgang.Schafer@ifj.edu.pl}
\affiliation{Institute of Nuclear Physics PAN, PL-31-342 Krak\'ow, Poland}

\author{Antoni~Szczurek}
\email{Antoni.Szczurek@ifj.edu.pl}
\affiliation{Institute of Nuclear Physics PAN, PL-31-342 Krak\'ow, Poland}
\affiliation{Institute of Physics, University of Rzesz\'ow, PL-35-959
  Rzesz\'ow, Poland}

\date{\today}%

\begin{abstract}
We discuss diffractive dissociation of gluons into heavy quark pairs.
The particular mechanism is similar to the diffractive dissociation
of virtual photons into quarks, which drives diffractive deep inelastic
production of charm in the low-mass diffraction, or large $\beta$-region.
There, it can be understood, with some reservations, in terms of a 
valence heavy quark content of the Pomeron.
The amplitude for the $g p \to Q \bar Q p$ is derived in the impact
parameter and momentum space. The cross section for single diffractive
$p p \to Q \bar Q p X$ is calculated as a convolution of the elementary
cross section and gluon distribution in the proton.
Integrated cross section and the differential distributions in e.g. transverse 
momentum and rapidity of the charm and bottom quark and antiquark, 
as well as the quark-antiquark invariant mass are calculated for the nominal LHC energy
for different unintegrated gluon distributions from the literature.
The ratio of the bottom-to-charm cross sections are shown and discussed
as a function of several kinematical variables. 
\end{abstract}

\maketitle

\section{Introduction}

Hard diffractive production is a special class of diffractive
processes. It is characterized by the production of massive objects
($W^{\pm}$, $Z^0$, Higgs boson, pairs of heavy quark - heavy antiquark)
or objects with large transverse momenta (jets, dijets) and one (single
diffractive process) or two (central diffractive process) rapidity gaps
between proton(s) and the centrally produced massive system.
The cross section for these processes is often calculated in terms of 
hard matrix elements for a given process and so-called diffractive
parton distributions, for a review, see e.g. \cite{Boonekamp:2009yd}.
The latter are often calculated, following a suggestion of Ingelman and Schlein
\cite{IS}, in a purely phenomenological approach in terms of parton distributions 
in the pomeron and a Regge-theory motivated flux of Pomerons. 
It is understood that such a factorization in hadron-hadron collisions must
be broken \cite{Bjorken:1992er}. To quantify absorptive corrections one often uses a global
gap survival factor.  For recent discussions on models of the gap survival,
see eg. \cite{KMR_eikonal,Maor}.

Diffractive production of heavy quarks was previously discussed 
within the Ingelman-Schlein model in
Refs.\cite{Fritzsch:1985wt,diffractive_open_charm_1,diffractive_open_charm_2,LMS2011}
and proposed as a probe of the hard subatructure of the Pomeron.

In this paper we wish to discuss a specific mechanism for the 
diffractive production of heavy quark -- antiquark pairs in 
proton-proton collisions in a ``microscopic approach'' which does not use the assumptions
of Regge factorization, and in which the QCD Pomeron is rather modelled by exchange of
a gluon ladder related to the unintegrated gluon distribution in the proton.

The mechanism we propose is based on the partonic subprocess $g p \to Q \bar Q p$ -- the
diffractive dissociation of a gluon into a heavy quark pair. At the first sight it
may be surprising that diffraction of a colored parton is a meaningful thing to consider,
but it turns out that the forward amplitude for the $g p \to Q \bar Q p$ is well
behaved and perturbatively calculable without introducing new soft parameters. 

Our mechanism has much in common -- in fact the both amplitudes turn out to be proportional to each other--
with the diffractive dissociation of a virtual photon into quarks $\gamma^* p \to Q \bar Q p$ 
through two-gluon exchange first introduced in \cite{Nikolaev:1991et}. 
Although there are many caveats, e.g. considering the Regge factorization, with certain reservations
the results of \cite{Nikolaev:1991et} can be formulated in terms of a valence quark structure 
of the Pomeron, see for example the discussions in \cite{Genovese:1994wx,Bartels:1998ea,Nikolaev:2004yh}.
The case of diffractive deep inelastic production of open charm was studied in this approach in detail 
first in \cite{Genovese:1995tc}. 

In the usual treatment of hard diffraction, heavy quarks are generated from gluons in the Pomeron
and a valence-like heavy quark contribution is not present. In this sense the mechanism discussed
here is complementary to existing approaches, although eventually we intend that the Ingelman-Schlein mechanism
be superseded also by a microscopic model for the glue in the Pomeron.

\begin{figure}[!h]
\includegraphics[width=6cm]{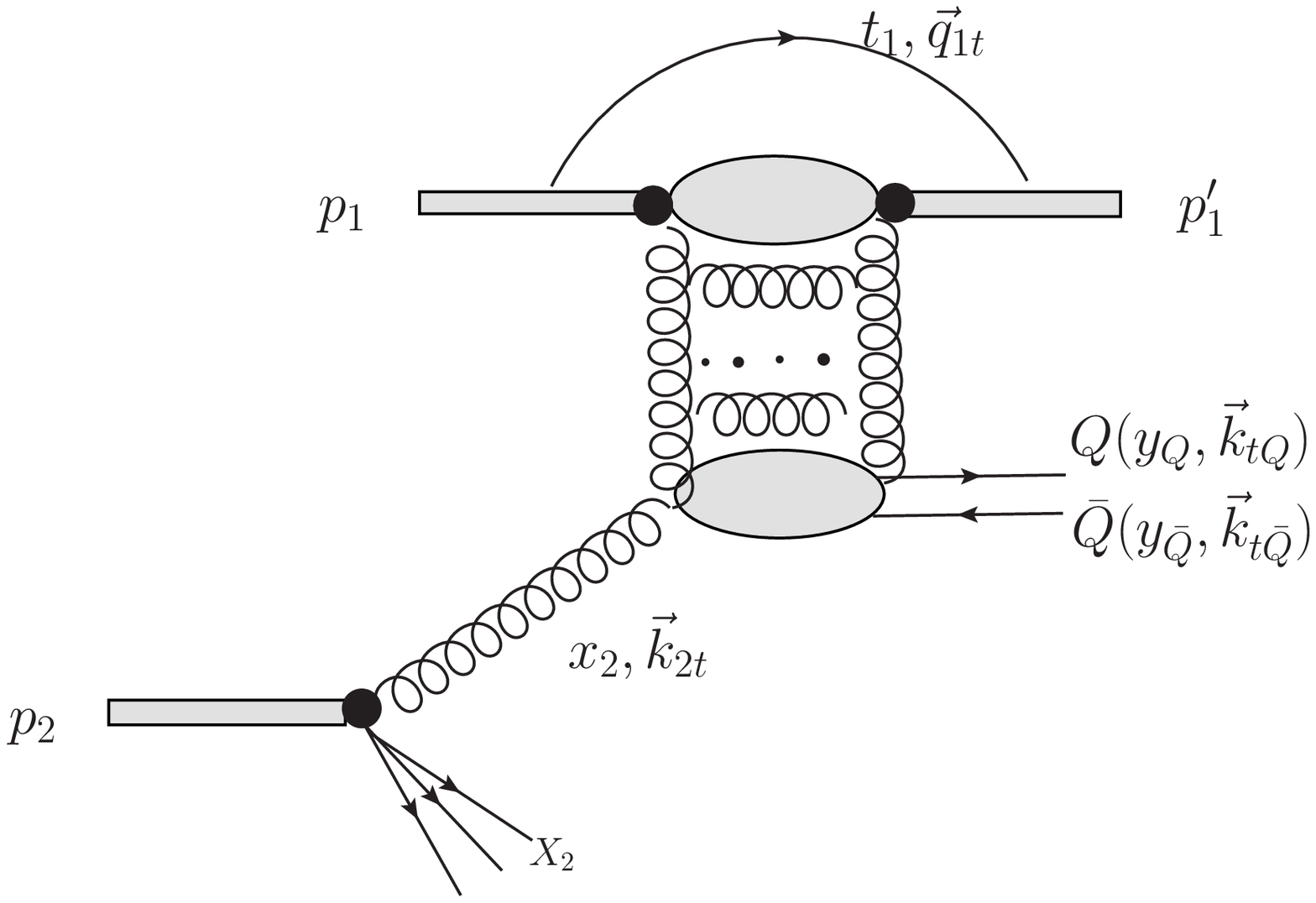}
\includegraphics[width=7cm]{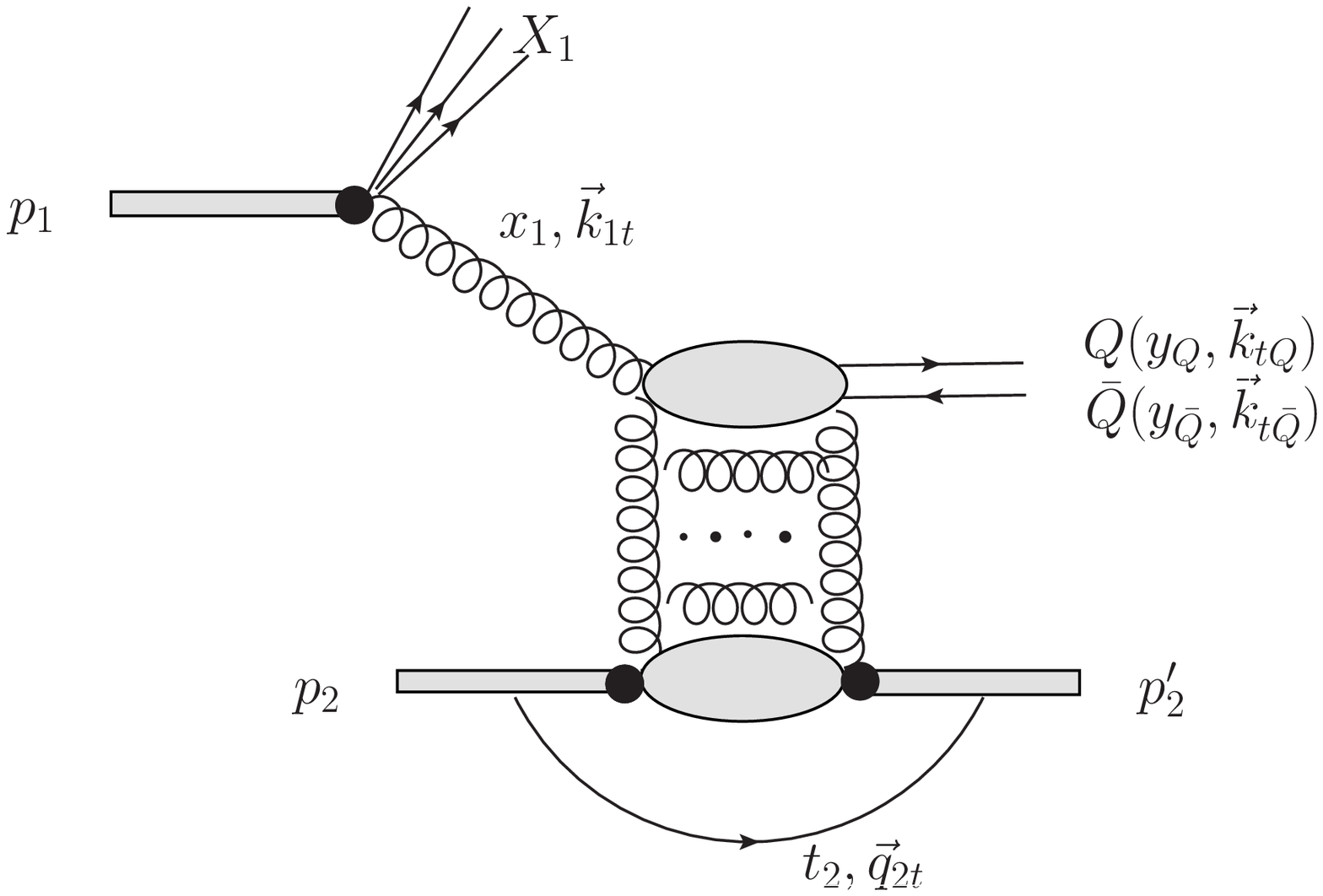}
   \caption{
\small The mechanism of gluon dissociation into $Q \bar Q$ via 
exchange of gluonic ladder in proton-proton collisions.
}
 \label{fig:sd_diagrams}
\end{figure}

From the experimental point of view there is a clear-cut kinematical distinction between 
the heavy quark production discussed of Fig.\ref{fig:sd_diagrams} here
and the standard approach: our $Q \bar Q$ pairs are produced 
in the Pomeron fragmentation region, close to the rapidity gap, whereas 
gluon-fusion generated heavy quarks populate a large part of the phase
space taken up by the diffractively produced system and will generally
give a tiny contribution in the Pomeron fragmentation region, unless there are a lot of 
hard gluons in the Pomeron.

In fact we are not the first to consider this mechanism of hard diffraction
in hadronic interactions: the same partonic amplitudes were calculated 
previously in \cite{Alves:1996ue,Yuan:1998ht}, although in an approximation
in which gluon transverse momenta in the Pomeron are integrated out.
Our results appear to be different from those presented in \cite{Alves:1996ue,Yuan:1998ht}.
We also mention that related, but different microscopic mechanisms are discussed in 
\cite{Kopeliovich:2007vs}), however only a Feynman $x_F$
distribution was presented there.

In the present paper we wish to present our amplitude for the
$g p \to Q \bar Q p$ (sub)process. We wish to calculate also
integrated and differential cross section for the $p p \to Q \bar Q p X$
single-diffractive processes at the LHC.
We will compare the cross section for the single diffractive process 
of charm and bottom.
Furthermore we will show some differential distributions, e.g. in
rapidity and transverse momentum for LHC energy.
Conclusions and outlook close our paper.

\section{Diffractive amplitude for $g p \to Q \bar Q p$}

A salient feature of high-energy interactions is that partons move along
straight line trajectories, and impact parameters are conserved during 
the interaction with the target, which is mediated by the $t$-channel
gluon exchanges.
\begin{figure}[!h]
\includegraphics[width=5cm]{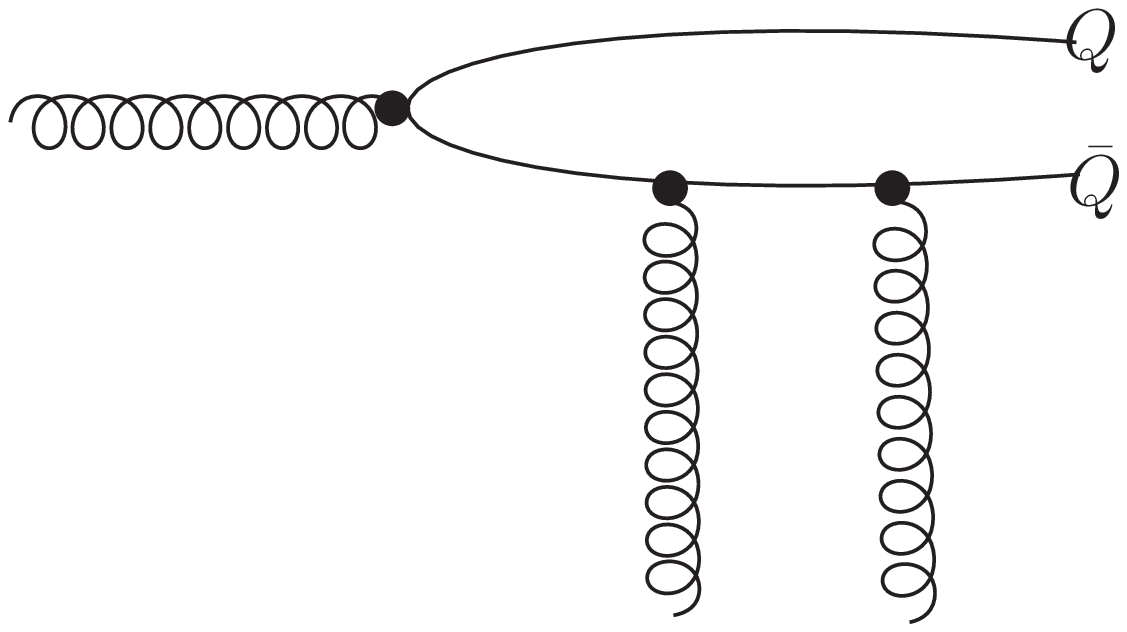}
\includegraphics[width=5cm]{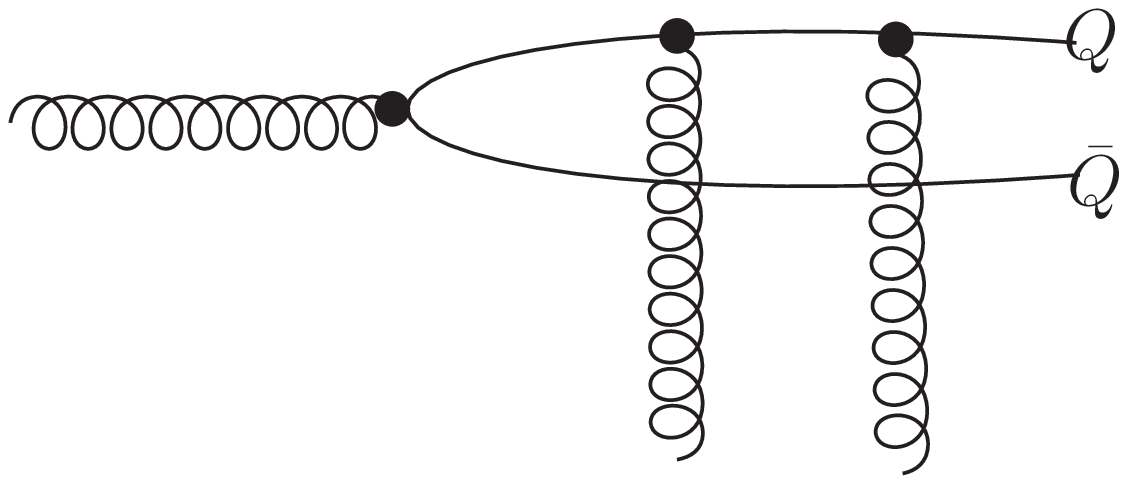}\\
\includegraphics[width=5cm]{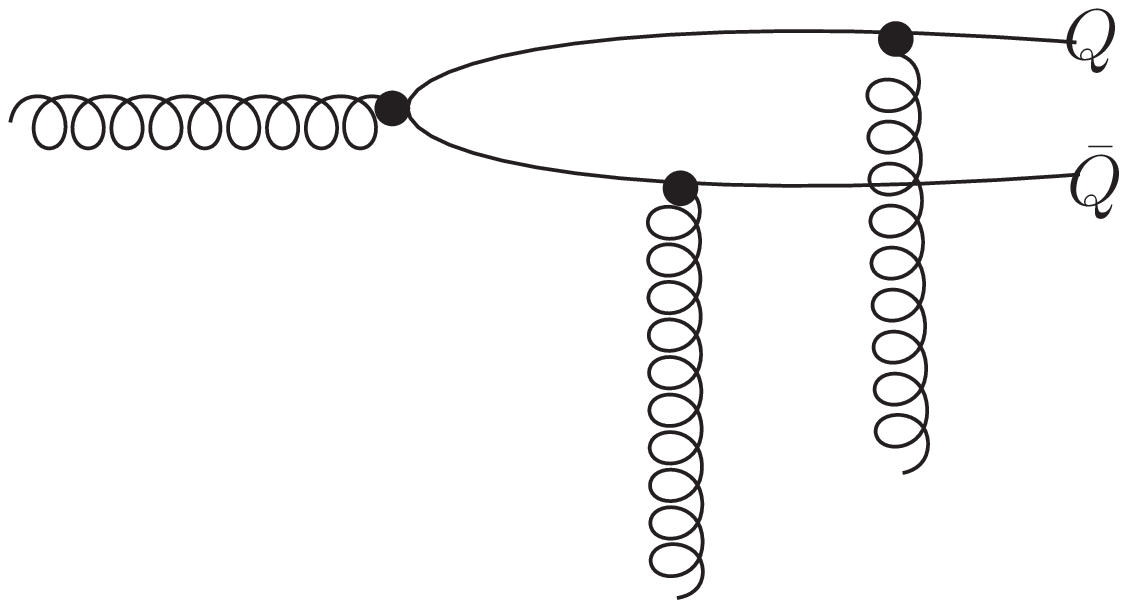}
\includegraphics[width=5cm]{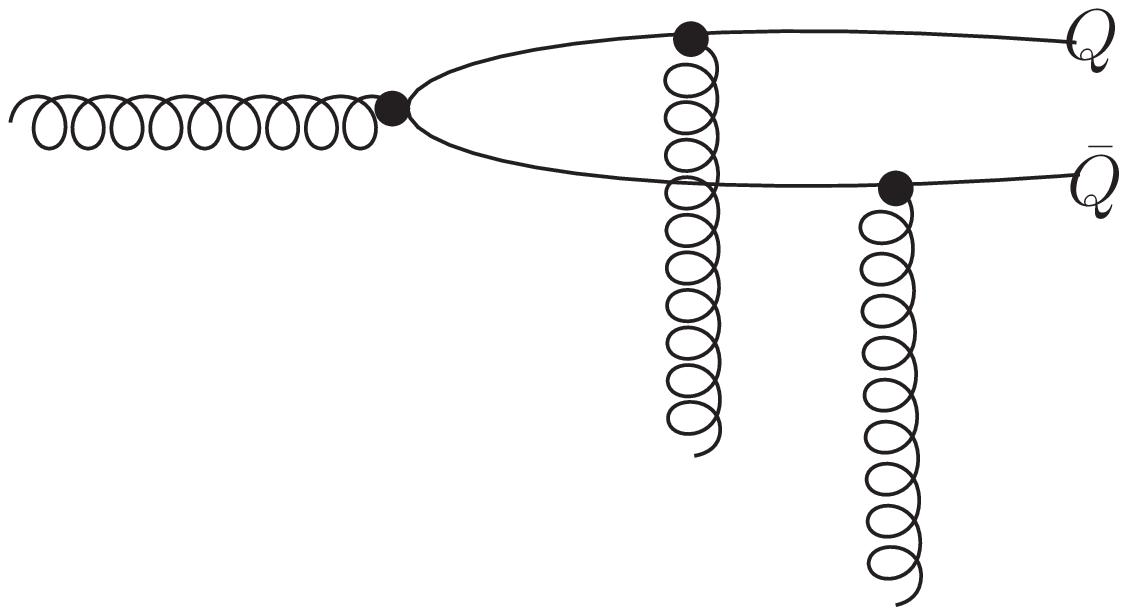}
\includegraphics[width=5cm]{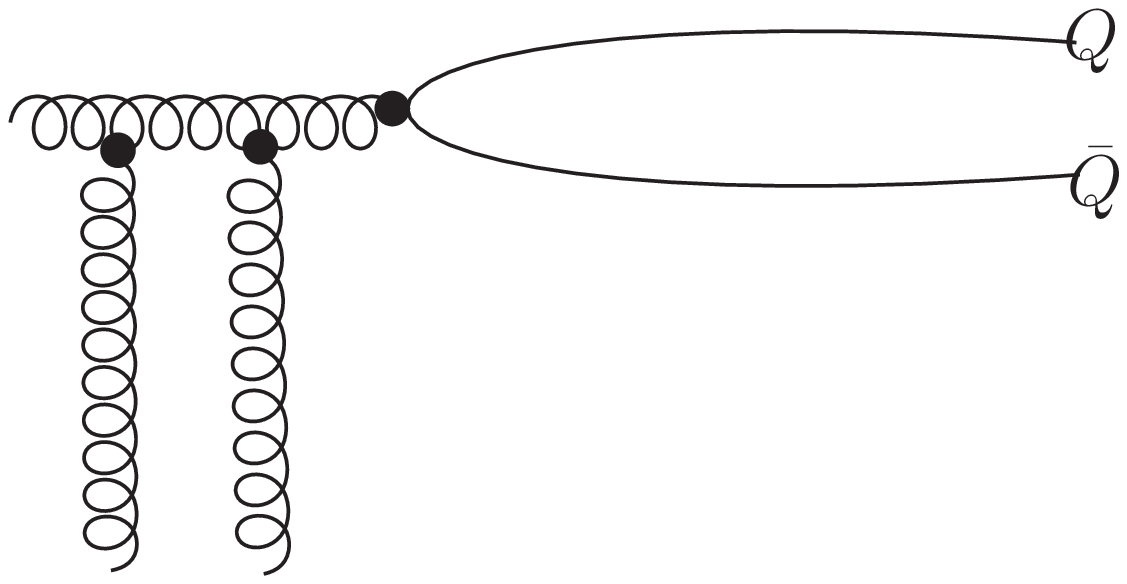}
   \caption{
\small Diagrams contributing to the diffractive
gluon dissociation into $Q \bar Q$. The two gluons in the $t$-channel 
are in the color singlet state. 
}
 \label{fig:diagrams_yes}
\end{figure}

\begin{figure}[!h]

\includegraphics[width=5cm]{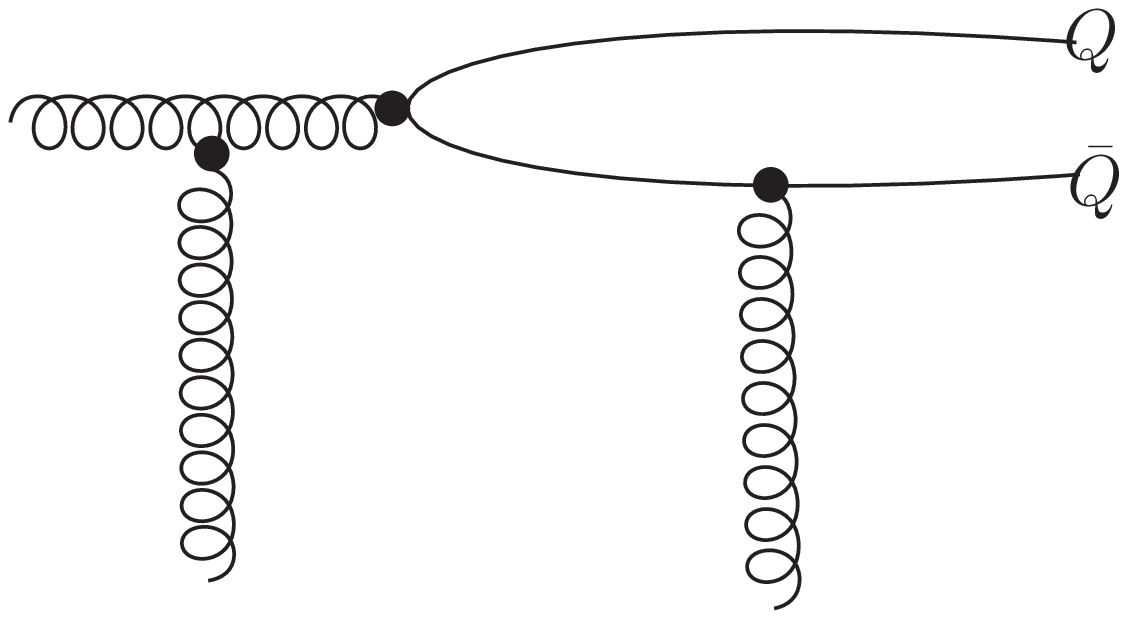}
\includegraphics[width=5cm]{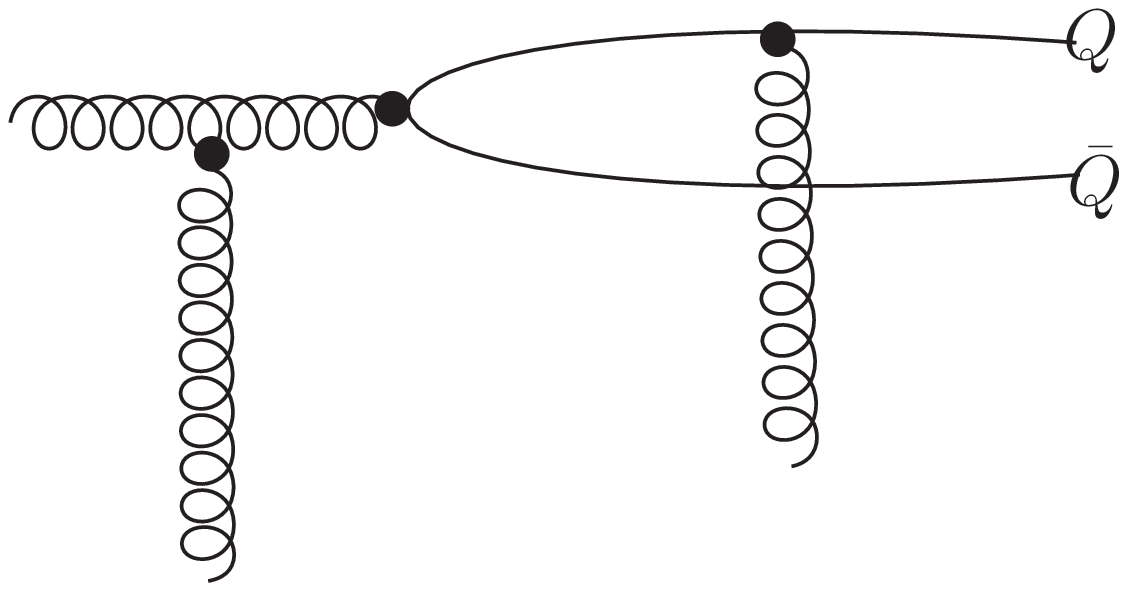}
   \caption{
\small Diagrams which do not contribute to the
gluon dissociation into $Q \bar Q$ in the high-energy limit.
}
 \label{fig:diagrams_no}
\end{figure}

The amplitude for the diffractive process $g N \to Q \bar Q N$ can be 
written in impact-parameter space, where the impact parameters of gluon,
quark and antiquark are $\bb,\bb_+,\bb_-$, respectively:

\begin{eqnarray}
{\cal{A}}_D(g_{\lambda_g}^a N \to Q_\lambda \bar Q_{\bar \lambda} N) = \int 
[{\cal{D}}\bb] \exp[-i\bp_+ \bb_+ - i \bp_- \bb_-] 
\bra{N} {\cal{M}}^a(\bb_+,\bb_-,\bb) \ket{N} \psi_{\lambda, \bar \lambda}^{\lambda_g} (z, \bb_+ - \bb_-) \, ,
\nonumber \\
\end{eqnarray}
The integration over impact parameters explicitly reads
\begin{eqnarray}
[{\cal{D}}\bb] =
d^2\bb_+ d^2\bb_- d^2\bb \, \delta^{(2)}(\bb - z \bb_+- (1-z) \bb_-) \, .  
\end{eqnarray}
Quark and antiquark share the lightcone-momentum of the incoming gluon 
in fractions $z,1-z$,  and have the transverse momenta $\bp_+, \bp_-$, respectively.
The transition $g_{\lambda_g} \to Q_\lambda \bar Q_{\bar \lambda}$ is described by the 
lightcone wave-function $\psi_{\lambda, \bar \lambda}^{\lambda_g}$. 
The interaction of partons with the target is encoded in the operator 
\cite{Nikolaev:2004cu} 
\begin{eqnarray}
{\cal{M}}^a(\bb_+,\bb_-,\bb) = 
\Big[ S(\bb_+) t^a S^\dagger(\bb_-) - S(\bb) t^a S^\dagger(\bb) \Big] \, .
\end{eqnarray}
Here the quark-proton $S$-matrix $S(\bb_+)$ is written in terms of the gluon-exchange
eikonal $\hat\chi = \chi^a t^a$, as
\begin{eqnarray}
S(\bb_+) = 1 + i \hat{\chi}(\bb_+) - {1 \over 2} \hat\chi^2(\bb_+) \, , 
\end{eqnarray}
the antiquark interacts with the complex-conjugate $S$-matrix $S^\dagger(\bb_-)$, and a gluon
interacts in the same fashion as a pointlike color-octet $Q\bar Q$ pair and its scattering
is described by $S(\bb) t^a S^\dagger(\bb)$ \cite{Nikolaev:2004cu}.

As we are interested in a diffractive process, with no color-transfer to
the target, in the evaluation of the nucleon-matrix element 
$\bra{N} {\cal{M}}^a(\bb_+,\bb_-,\bb) \ket{N}$,
only the terms of quadratic order in the eikonal will contribute. 
Indeed, they correspond to
the coupling of two gluons in the color-singlet state to the nucleon:
\begin{eqnarray}
\bra{N} \chi^a(\bb_+) \chi^b(\bb_-) \ket{N} = \delta^{ab} \, \chi^{(2)}(\bb_+,\bb_-) \, ,
\end{eqnarray}
which we parametrize as
\begin{eqnarray}
C_F \cdot \chi^{(2)}(\bb_+,\bb_-) = \int {d^2\bq \over (2 \pi)^2} {d^2 \bkappa \over (2 \pi)^2}  
\, f(x, {\bq \over 2} + \bkappa, {\bq \over 2} - \bkappa) \, \exp[i{\bq \over 2} (\bb_+ + \bb_-)]
\, \exp[i \bkappa(\bb_+ - \bb_-)] \, .
\nonumber \\
\end{eqnarray}
Here the function $f$ is a generalized two-gluon density matrix, including 
gluon propagators, and off-diagonal in transverse momenta. It takes the form
\begin{eqnarray}
 f(x, \bkappa_1, \bkappa_2) = { (2 \pi)^3 \over N_c} \cdot
{\alpha_S \, {\cal{F}}(x, \bkappa_1, \bkappa_2) \over 
\bkappa_1^2 \bkappa_2^2} \, , 
\end{eqnarray}
where in turn ${\cal{F}}$, at large $\bkappa^2$ is related to the familiar unintegrated gluon 
structure function as
\begin{eqnarray}
{\cal{F}}(x,  \bkappa, - \bkappa) = 
{\partial [xg(x,\bkappa^2)] \over \partial \log(\bkappa^2)} \, . 
\end{eqnarray}
The full $f$ is often parametrized as
\begin{eqnarray}
  f(x, {\bq \over 2} + \bkappa, {\bq \over 2} - \bkappa) = {(2 \pi)^3 \over N_c} 
{\alpha_S \over \bkappa^4} {\partial [xg(x,\bkappa^2)] 
\over \partial \log(\bkappa^2)} \,
\exp[-{1 \over 2} B_D \bq^2] \, ,
\end{eqnarray}
where the diffractive slope $B_D$ is a nonperturbative quantity that
takes care of the effective size of the target.
In accord with Regge-phenomenology, it has an $x$-dependent piece, 
$B_D = B_0 + \alpha'_\Pom \log(x_0/x)$. In our numerical calculations, we use an educated guess
for the diffractive slope \cite{Ivanov:2004ax}: $B_0 = 4.0 \, \mathrm{GeV}^{-2}, \alpha'_\Pom = 0.164$.

We then obtain:
\begin{eqnarray}
 \bra{N} {\cal{M}}^a(\bb_+,\bb_-,\bb) \ket{N} =  t^a \Big\{ 
{1 \over 2 N_c C_F} \, \Gamma^{(2)}(\bb_+,\bb_-)
-{N_c \over 2 C_F}  \, \bar{\Gamma}^{(2)}(\bb_+,\bb_-) \Big\} \, ,
\end{eqnarray}
where 
\begin{eqnarray}
 \Gamma^{(2)}(\bb_+,\bb_-) &=& {C_F \over 2} \cdot \Big[
\chi^{(2)}(\bb_+,\bb_+)
+\chi^{(2)}(\bb_-,\bb_-)  
-\chi^{(2)}(\bb_+,\bb_-)
-\chi^{(2)}(\bb_-,\bb_+)
\Big]  \; ,
\nonumber \\
\bar{\Gamma}^{(2)}(\bb_+,\bb_-) &=&
{C_F \over 2} \cdot \Big[
\chi^{(2)}(\bb_+,\bb_+)
+\chi^{(2)}(\bb_-,\bb_-)  
-2 \chi^{(2)}(\bb,\bb)
\Big] \, .
\end{eqnarray}
Here the profile function $\Gamma^{(2)}(\bb_+,\bb_-)$ is related to 
the familar color-dipole cross section through the relation
\begin{eqnarray}
\sigma(\br) = 2 \, \int d^2\bB \, \Gamma^{(2)}(\bB+{\br \over 2}, \bB - {\br \over 2}) \, . 
\end{eqnarray}
Following \cite{Ivanov:2004dg}, we can write the relevant profile functions in terms of the off-diagonal
gluon density as:
\begin{eqnarray}
\Gamma^{(2)}(\bb_+,\bb_-) &=& \int {d^2\bq \over (2 \pi)^2} {d^2 \bkappa \over (2 \pi)^2}  
\, f(x, {\bq \over 2} + \bkappa, {\bq \over 2} - \bkappa) 
\, \exp[i {\bq \over 2}(\bb_+ +\bb_-)] \, 
\nonumber \; , \\
&&
\times \Big\{
\exp[i {\bq \over 2} (\bb_+-\bb_-)] 
+ \exp[-i{\bq \over 2}(\bb_+-\bb_-)]
\nonumber \\
&&- \exp[i \bkappa(\bb_+ - \bb_-)] 
- \exp[-i \bkappa(\bb_+ - \bb_-)]
\Big\} \, , \nonumber \\
\bar{\Gamma}^{(2)}(\bb_+,\bb_-) &=& \int {d^2\bq \over (2 \pi)^2} {d^2 \bkappa \over (2 \pi)^2}  
\, f(x, {\bq \over 2} + \bkappa, {\bq \over 2} - \bkappa) 
\, \Big[ \exp[i \bq\bb_+] + \exp[i \bq\bb_-] -2 \exp[i \bq\bb] \Big] \, .
\nonumber \\
\end{eqnarray}
We now introduce the usual parametrization of transverse momenta: 
the decorrelation momentum of jets (or momentum transfer to the proton) 
$\bDelta = \bp_+ + \bp_-$ and the light-cone relative transverse 
momentum $\bk = (1-z) \bp_+ - z \bp_-$,
which is conjugate to the dipole size $\br = \bb_+ - \bb_-$.
We then notice, that
\begin{eqnarray}
 [{\cal{D}}\bb] \exp[-i\bk (\bb_+ -\bb_-) - i \bDelta(z \bb_+ + (1-z) \bb_-)] = 
 d^2\bb d^2\br \, \exp[-i \bDelta \bb] \, \exp[-i\bk \br] \, , 
\end{eqnarray}
so that
\begin{eqnarray}
{\cal{A}}_D(g^a N \to Q \bar Q N) &&= 
\int d^2\bb d^2\br \, \exp[-i \bDelta \bb] \exp[-i\bk \br] \, 
\psi_{\lambda, \bar \lambda}^{\lambda_g} (z, \br) 
\nonumber \\
&& 
\times \bra{N} {\cal{M}}^a(\bb+(1-z)\br,\bb-z\br,\bb) \ket{N} \nonumber\\
&&= t^a \, 
\int d^2\bb d^2\br \, \exp[-i \bDelta \bb]\exp[-i\bk \br] \, 
\psi_{\lambda, \bar \lambda}^{\lambda_g} (z, \br) \nonumber\\
&&
\times \Big\{ 
{1 \over 2 N_c C_F} \, \Gamma^{(2)}(\bb+(1-z)\br,\bb-z\br) 
-{N_c \over 2 C_F}  \, \bar{\Gamma}^{(2)}(\bb+(1-z)\br,\bb-z\br) \Big\} \, . \nonumber\\
\end{eqnarray}
Now, the term $\propto \bar \Gamma^{(2)}$ vanishes in the forward
direction $\bDelta = 0$.
For the forward amplitude we can then easily derive:
\begin{eqnarray}
 {\cal{A}}_D(g^a N \to Q \bar Q N)|_{\bDelta =0} &=& 
{t^a \over 2 N_c C_F}  {4 \pi \over N_c} \nonumber \\
&&  \times \int {d^2\bkappa \over \bkappa^4} \, \alpha_S {\cal{F}}(x_{\mathrm{eff}}, \bkappa,-\bkappa) 
\Big[ \psi_{\lambda, \bar \lambda}^{\lambda_g} (z, \bk) - \psi_{\lambda,
  \bar \lambda}^{\lambda_g} (z, \bk+\bkappa) \Big] \; .
\nonumber \\
\end{eqnarray}
Here we have introduced the argument $x_{\mathrm{eff}}$ of the unintegrated gluon distribution,
for which we take:
\begin{equation}
x_{\mathrm{eff}} = x_\Pom = {M^2 \over \hat{s}}, \mathrm{with} \,  M^2 = {\bk^2 + m_Q^2 \over z (1-z)} \, .
\end{equation}
As a matter of principle we should use an unintegrated gluon
distribution which is also off-diagonal in the longitudinal momentum
fractions of gluons. Here we neglect the effects of ``skewedness'',
which could be expected to enhance the cross section. Remember that in the hadron--level 
cross section there has a large uncertainty regarding the gap survival probability. In this view, 
we believe that our neglect of skewedness, and educated guessing of the slope, is justified.
Furthermore various ratios shown in the results section should not be overly sensitive to these
somewhat crude approximations.

Our amplitude can now be expressed in terms of the integrals that are 
found in \cite{Nikolaev:1998wf}:
\begin{equation}
  \left.\begin{aligned}
         \bPhi_1& \\
         \Phi_0&
        \end{aligned}
  \right\}
   = \int_0^\infty {d\bkappa^2 \over \bkappa^4}  \alpha_S {\cal{F}}(x_{\mathrm{eff}}, \bkappa,-\bkappa)  
\begin{cases}
\displaystyle
{\bk \over \bk^2 + m_Q^2} \, W_1(\bk^2,\bkappa^2,m_Q^2)  \\
\displaystyle
{1 \over \bk^2 + m_Q^2} W_0(\bk^2,\bkappa^2,m_Q^2) \, ,
\end{cases}
\end{equation}
\begin{eqnarray}
W_1(\bk,\bkappa^2,m_Q^2) &=&  1 -  \theta(\bk^2,\bkappa^2,m_Q^2)   \; , \\
W_0(\bk,\bkappa^2,m_Q^2) &=&  1 - {\bk^2 + m_Q^2 \over  \sqrt{(\bk^2 -
    m_Q^2 - \bkappa^2)^2 + 4 \bk^2 m_Q^2}} \; ,
\end{eqnarray}
and 
\begin{eqnarray}
\theta(\bk^2,\bkappa^2,m_Q^2) = {\bk^2 + m_Q^2 \over 2 \bk^2} \, 
\Big( 1 + {\bk^2 -m_Q^2 - \bkappa^2 \over  \sqrt{(\bk^2 - m_Q^2 - \bkappa^2)^2 + 4 \bk^2 m_Q^2}}\Big) \, .
\end{eqnarray}
Let us write out the amplitude explicitly for the different helicities (see e.g Eq.(108) in \cite{Ivanov:2004ax},
here $\be(\lambda_g) = -(\lambda_g \be_x + i \be_y)/\sqrt{2}$):
\begin{eqnarray}
{\cal{A}}_D(g_{\lambda_g}^a N \to Q_\lambda \bar Q_{\bar \lambda} N)|_{\bDelta =0}
&=& {t^a \over 2 N_c C_F} {4 \pi^2 \over N_c} \sqrt{\alpha_S} \nonumber \\
&&
\Big\{
2 \delta_{\lambda -\bar \lambda} 
\Big[ 
z \delta_{\lambda_g \lambda} - (1-z) \delta_{\lambda_g \bar \lambda}
\Big] (\bPhi_1 \cdot \be(\lambda_g))
\nonumber \\
&& - \delta_{\lambda \bar \lambda} \delta_{\lambda_g \lambda} \sqrt{2} m_Q \Phi_0 
\Big\}
\end{eqnarray}
Then, if we put everything together, we get for the differential 
parton-level cross section.
\begin{eqnarray}
16 \pi {d\hat \sigma(g N \to Q\bar Q N;\hat{s}) \over d\bDelta^2}\Big|_{\bDelta^2=0} = {1 \over 2 \cdot (N_c^2-1)} \cdot
\sum_{\lambda_g,\lambda,\bar\lambda,a} \Big| {\cal{A}}_D(g_{\lambda_g}^a N \to Q_\lambda \bar Q_{\bar \lambda} N) \Big|^2 \, dz 
{d^2\bk \over (2 \pi)^2} \, , 
\nonumber \\
\end{eqnarray}
and the final multi-dimensional cross section reads:
\begin{eqnarray}
{d \hat \sigma (gN \to Q \bar Q N;\hat{s}) \over dz d^2\bk d\bDelta^2}\Big|_{\bDelta^2=0} = {\pi  \over 4 \, N_c^2 (N_c^2-1)^2} \,\alpha_S 
\Big \{ [z^2 + (1-z)^2] \bPhi_1^2 + m_Q^2 \Phi_0^2 \Big\}  \, .
\label{eq:parton-level}
\end{eqnarray}

\subsection{Hard scale, quark-mass dependence and twist-expansion}
In order to expose the dependence on the heavy quark mass $m_Q$, and/or
the hard scale of the process $\bar Q^2 = \bk^2 + m_Q^2$, it is useful to develop the twist
expansion of the diffractive amplitude and extract the piece leading in $\bar Q^2$.

The starting point would be an expansion of weight functions $W_{0,1}$ in powers of
$\bkappa^2/\bar Q^2$. From the Taylor expansion
\begin{eqnarray}
{1 \over  \sqrt{(\bk^2 - m_Q^2 - \bkappa^2)^2 + 4 \bk^2 m_Q^2}} =
{ 1\over \bar Q^2} \Big( 1 + \bkappa^2 {\bk^2 - m_Q^2 \, \over \bar{Q}^4}  + \dots \Big) \, ,
\end{eqnarray}
we obtain
\begin{eqnarray}
W_1(\bk,\bkappa^2,m_Q^2) \simeq  {2 m_Q^2 \over [\bk^2 + m_Q^2]^2} \cdot
\bkappa^2 \, , \, 
W_0(\bk,\bkappa^2,m_Q^2) \simeq {\bk^2 - m_Q^2 \over [\bk^2 + m_Q^2]^2} \cdot \bkappa^2 \, .
\end{eqnarray}
As it must be, the weigth functions for small $\bkappa^2$ vanish 
$\propto \bkappa^2$, so that we obtain
a large contribution to the diffractive amplitude which is proportional to the integrated
gluon distribution $xg(x,\bar Q^2)$:
\begin{eqnarray}
\int^{\bar Q^2} {d \bkappa^2 \over \bkappa^4} \cdot \bkappa^2 \cdot \alpha_S {\partial [xg(x,\bkappa^2)] \over \partial \log(\bkappa^2)}
 = \alpha_S(\bar Q^2 ) xg(x,\bar Q^2) \, ,
\end{eqnarray}
where $\bar Q^2 = \bk^2 + m_Q^2$. This is the ``leading twist''
contribution in which virtualities $\bkappa^2$ of exchanged gluons are much
smaller than the virtualities of the quark lines, and thus we can integrate 
out the gluon transverse momenta.
Introducing $\tau \equiv \bk^2/m_Q^2$, we find for the differential cross section
(\ref{eq:parton-level}) the result
\begin{eqnarray}
{d \hat \sigma (gN \to Q \bar Q N;\hat{s}) \over dz d\tau d\bDelta^2}\Big|_{\bDelta^2=0} &=& {\pi^2  \over 4 \, N_c^2 (N_c^2-1)^2} { 1 \over m_Q^4} \, 
{ (1 - \tau)^2 + [z^2 + (1-z)^2] 4 \tau \over (1 + \tau)^6}  \nonumber \\
&&\times \alpha_S^3( (1 + \tau) m_Q^2) \Big[ x_{\mathrm{eff}} g (x_\mathrm{eff}, (1 + \tau) m_Q^2 ) \Big]^2 \, .
\label{eq:approx}
\end{eqnarray}
Here the scaling of the cross section with $m_Q^{-4}$ is put in evidence. It is only violated 
by the argument of the running coupling and the integrated glue of the target. 
We notice that this approximation is used throughout in earlier works on charm production 
in deep inelastic and hadronic reactions \cite{Genovese:1995tc,Alves:1996ue,Yuan:1998ht}.

We note parenthetically, that the leading twist approximation is expected to work well for $\tau \lsim 1$ and
for the integrated cross section. It will be inadequate for jet production, at large jet transverse momenta $\tau \gg 1$. 
In this case the weight functions $W_{0,1}$ will develop sharp peaks at around 
$\bkappa^2 \sim \bk^2$, the twist expansion breaks down, and 
the diffractive amplitude will then be dominated by a piece 
that is directly proportional to the unintegrated gluon distribution 
of the proton \cite{Nikolaev:1994cd}.

We go beyond the approximation (\ref{eq:approx}), and in our numerical calculations always use the
$\bk_\perp$-factorization representation (\ref{eq:parton-level}).

Finally a comment on earlier works is in order. Alves et al. \cite{Alves:1996ue} replace in 
the diagrams of Fig.\ref{fig:diagrams_yes} the incoming gluon by an effective color singlet current, and hence
have no diagram in which the two $t$-channel gluons couple to the incoming gluon. While
this gives the correct structure of the amplitude, the correct normalization can only be
found by doing the proper color algebra in the full set of diagrams. Furthermore, Alves at al. \cite{Alves:1996ue}
also consider diagrams of the type shown in Fig.\ref{fig:diagrams_no}. Notice that when the two $t$-channel gluons are coupled
to the proton, each of these diagrams has two cuts which contribute to the imaginary part of the amplitude, one in which the
$s$-channel gluon line is cut, and another one where the (anti-)quark line is cut. In the high energy limit of
$M_{Q \bar Q}^2/\hat s \ll 1$ these two contributions cancel each other. This is a technical manifestation 
of an ``LPM''-like phenomenon: in the limit of large coherence length, the radiation between scatterings vanishes
(for a review of LPM-physics in perturbative QCD, see e.g. \cite{Baier:2000mf}). In Ref.\cite{Alves:1996ue} the diagrams
of Fig.\ref{fig:diagrams_no} give the dominant contribution, but it appears that only the cut through the gluon line
has been accounted for, and the cut through the (anti-)quark line apparently has been overlooked.

The authors of Ref.\cite{Yuan:1998ht} state, that the diagrams of Fig.\ref{fig:diagrams_no} are necessary to
cancel soft divergences -- this is not borne out by our calculation, our
final results also differ from those in \cite{Yuan:1998ht}.

\subsection{Hadron-level cross section}
\label{subsection:hadron_level}
We finally want to calculate for example the spectrum of quarks in 
the $pp$-collision. Starting from the diffractive $g p \to Q \bar Q p$ cross section
\begin{eqnarray}
 {d \hat \sigma (gN \to Q \bar Q N;\hat{s}) \over dz d^2\bk d\bDelta^2}\Big|_{\bDelta^2=0} = \hat f_{Q \bar Q}(z,\bk;\hat s) \, 
\end{eqnarray}
we can obtain the corresponding cross section for $pp$-collisions in the collinear approximation for the incoming gluon
as:
\begin{eqnarray}
{d \sigma (pp \to X Q\bar Q + p;s) \over dx_Q d^2\bk d\bDelta^2}\Big|_{\bDelta^2=0} &=&
\int dx dz \, \delta(x_Q - xz) \, g(x,\bar Q^2) \,  \hat f_{Q \bar Q}(z,\bk;xs) \, 
\; ,
\nonumber \\
&=& \int_{x_Q}^1  {dx \over x} g(x,\bar Q^2) \,  \hat f_{Q \bar Q} \Big( {x_Q \over x}, \bk; x s \Big) \, .
\end{eqnarray}

We can also calculate the  fully differential distribution in $x_Q, x_{\bar Q} = x - x_Q$, rapidities etc.
\begin{eqnarray}
{d \sigma (pp \to X Q\bar Q + p;s) \over dx_Q dx_{\bar Q} d^2\bk d\bDelta^2}\Big|_{\bDelta^2=0} &=&
{1 \over x_Q + x_{\bar Q}}  \, g(x_Q + x_{\bar Q},\bar Q^2) \,  \hat f_{Q \bar Q} \Big( {x_Q \over x_Q + x_{\bar Q}},\bk;xs \Big) \, .
\end{eqnarray}
Alternatively, the rapidity distributions of quarks may be of interest.
Let the incoming protons have the four momenta 
\begin{eqnarray}
p_A = \sqrt{s \over 2} n_+  + {m_p^2 \over 2}\sqrt{2 \over s} n_- \, , \, p_B = {m_p^2 \over 2}\sqrt{2 \over s} n_+ + \sqrt{s \over 2} n_- \, ,  
\end{eqnarray}
with
\begin{eqnarray}
n_\pm = {1 \over \sqrt{2}} (1,0,0,\pm 1) \, , p_A^2 = p_B^2 = m_p^2 \, ,\, 2 p_A p_B = s + {\cal{O}}({m^4 \over s}) .
\end{eqnarray}
The four-momenta of quark and antiquark are then
\begin{eqnarray}
p_Q &=& x_Q p_A^+ \, n_+ + {\bk^2 + m_Q^2 \over 2 x_Q p_A^+} \, n_ - +
k_\perp \;, \nonumber \\
p_{\bar Q} &=& x_{\bar{Q}} p_A^+ \, n_+ + {\bk^2 + m_Q^2 \over 2 x_{\bar Q} p_A^+} \, n_- - k_\perp \; .
\end{eqnarray}
Instead of $x_Q, x_{\bar Q}$ one often uses the rapidities:
\begin{eqnarray}
y_Q = {1 \over 2} \log \Big( {p_Q^+ \over p_Q^-} \Big) \, , \,
y_{\bar{Q}} = {1 \over 2} \log \Big( {p_{\bar Q}^+ \over p_{\bar Q}^-}
\Big) \; .
\end{eqnarray}
Explicitly:
\begin{eqnarray}
y_Q = \log \Big( {x_Q \sqrt{s} \over \sqrt{\bk^2 + m_Q^2}} \Big) \, , \, y_{\bar Q} = \log \Big( {x_{\bar Q} \sqrt{s} \over \sqrt{\bk^2 + m_Q^2} }\Big) \, .
\end{eqnarray}
The rapidity difference $\Delta y$, and average rapidity $Y = (y_Q + y_{\bar Q} )/2$  of quark and antiquark are
\begin{eqnarray}
\Delta y &=& y_Q - y_{\bar Q} = \log \Big( {x_Q \over x_{\bar Q} } \Big) = \log \Big( {z \over 1-z} \Big ) \, , \nonumber \\
Y &=& {1 \over 2} (y_Q + y_{\bar Q}) = {1 \over 2} \log  \Big( {x_Q  x_{\bar Q}  s \over \bk^2 + m_Q^2 } \Big) = { 1\over 2} \log \Big( { x \over x_\Pom} \Big) \, .
\end{eqnarray}
Notice that the dissociating proton is at positive rapidity.
In fact $y_A = \log ( \sqrt{s}/m_p ) \sim + 9.5$, while $y_B =  \log
(m_p / \sqrt{s} ) \sim - 9.5$ at the nominal LHC energy.

\section{Numerical results}
In our numerical calculations, we shall use
three different unintegrated gluon distribution functions (UGDFs) from the literature.
One of them from Ref.\cite{IN2002} (labelled Ivanov-Nikolaev) is a fit
to HERA structure functions data.
The other two, from Ref.\cite{KS2005} (labelled Kutak-Stasto) are
obtained by solving a BFKL equation accounting for subleading terms. One of the latter UGDFs also accounts for a 
nonlinear Balitsky-Kovchegov type term in the evolution equation. 
Both UGDF sets give a reasonable description of deep inelastic
structure functions at small $x$.

For the argument of running coupling constant we take: 
$\mu_r^2 = M_{Q \bar Q}^2$ for $g \to Q \bar Q$ splitting and
$\mu^2 = \max(\kappa^2, \bk^2+m_Q^2)$ for the $t$-channel coupling of gluons
to heavy quarks. 
For the quark masses, we take $m_c = 1.5 \, \mathrm{GeV}$ and
$m_b = 4.75 \, \mathrm{GeV}$

Before we shall show differential distributions let us discuss integrated
cross section for single diffractive process for $Q \bar Q$ production.
In Table \ref{tab:cross_section} we have collected the integrated cross
section for different UGDFs and different GDFs from the literature
\cite{GRV94,CTEQ6,GJR}.
Here we have not included a gap survival factor.
The cross section for $b \bar b$ is about two orders of magnitude smaller
than that for $c \bar c$, which is well in line with the expected $m_Q^{-4}$ scaling:
$(m_c/m_b)^4 \sim 1 \%$. 
We observe a dependence of the cross
section on the choice of UGDF, especially for $c \bar c$ production. 
Particularly interesting is the difference between results with 
linear and nonlinear versions of the Kutak-Sta\'sto UGDF. 
The difference between different UGDFs points to the fact that single diffractive production 
of charm tests UGDFs in the region of very small $x_\Pom$. We shall return 
to the impact of UGDFs on differential distributions somewhat later in 
this section.
The dependence of the cross section on the choice
of different GDFs is of a similar size. 
Again the largest differences are observed for charm production, 
which is expected as the small-$x$ glue is probed at somewhat smallish
hard scales. 
\begin{table}[tb]
\caption{Integrated cross section in mb for single diffractive
production of $c \bar c$ and $b \bar b$ for different UGDFs and GDFs
at $\sqrt{s}$ = 14 TeV. No gap survival factor was included here.}
\label{tab:cross_section}
\begin{tabular}{|c|c|c|c|c|}
\hline
 UGDF        &  GDF      &  $\sigma(c \bar c)$ [mb] & $\sigma(b \bar b)$ [mb]  & $\sigma( b \bar b)/\sigma(c \bar c) $\\
\hline
IN            &  GRV94    &  1.3891 e-2     &  1.4201 e-4 &  1 \% \\
              &  GJR08    &  1.0954 e-2     &  1.1083 e-4 &  1 \% \\
              &  CTEQ6    &  1.7607 e-2     &  1.3808 e-4 &  0.7 \% \\
\hline
KS (lin.)     &  GRV94    &  1.4837 e-2     &  1.0293 e-4 &  0.7 \% \\
              &  GJR08    &  1.0402 e-2     &  0.9011 e-4 &  0.9 \%\\
              &  CTEQ6    &  1.9719 e-2     &  1.0789 e-4 &  0.5 \%\\
\hline
KS (non lin.) &  GRV94    &  1.0122 e-2     &  0.9185 e-4 &  0.9 \% \\
              &  GJR08    &  0.8322 e-2     &  0.9005 e-4 &  1 \% \\
              &  CTEQ6    &  1.2335 e-2     &  0.9311 e-4 &  0.8 \%  \\
\hline
\end{tabular}
\end{table}

\begin{figure}[!h]
\includegraphics[width=7cm]{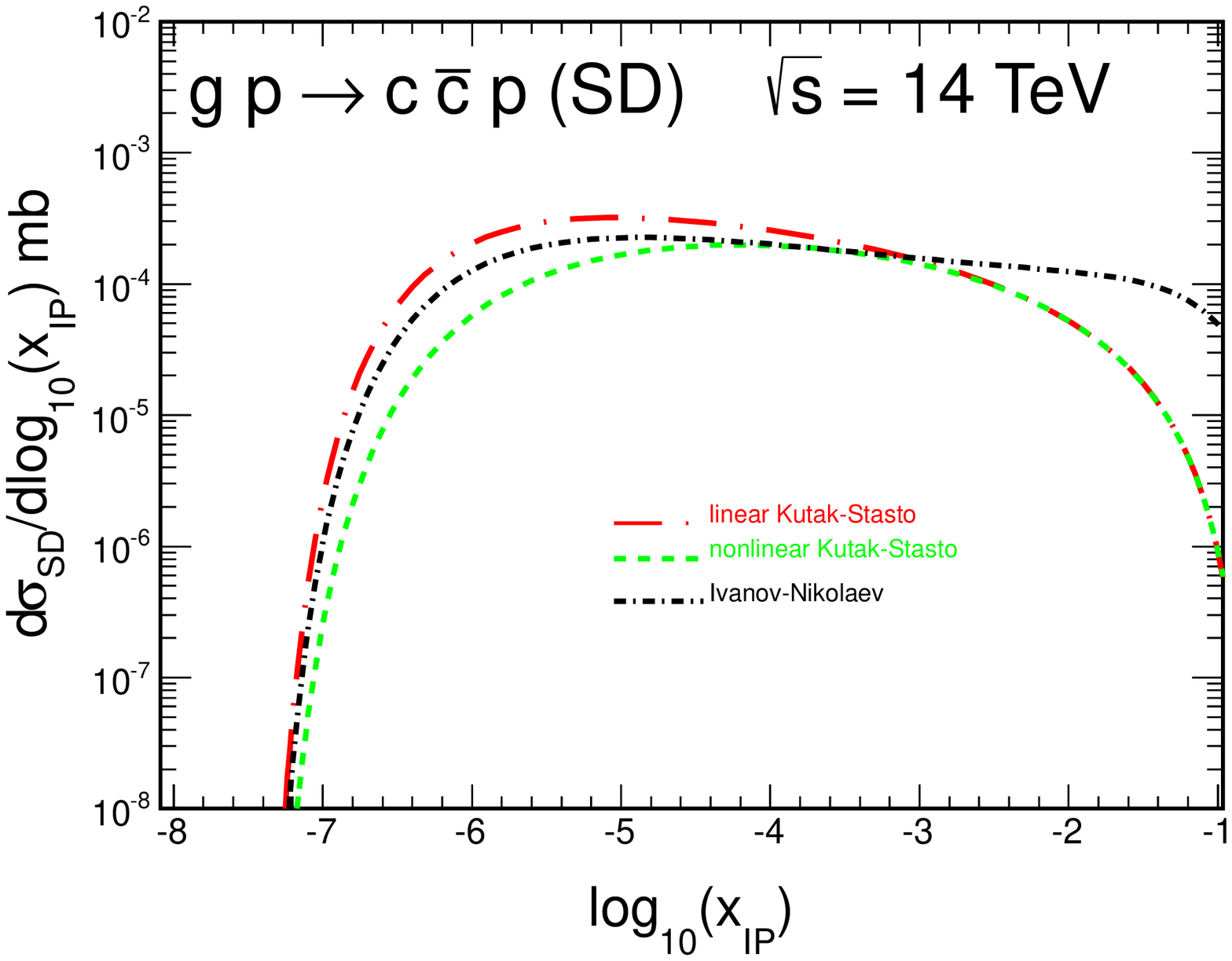}
\includegraphics[width=7cm]{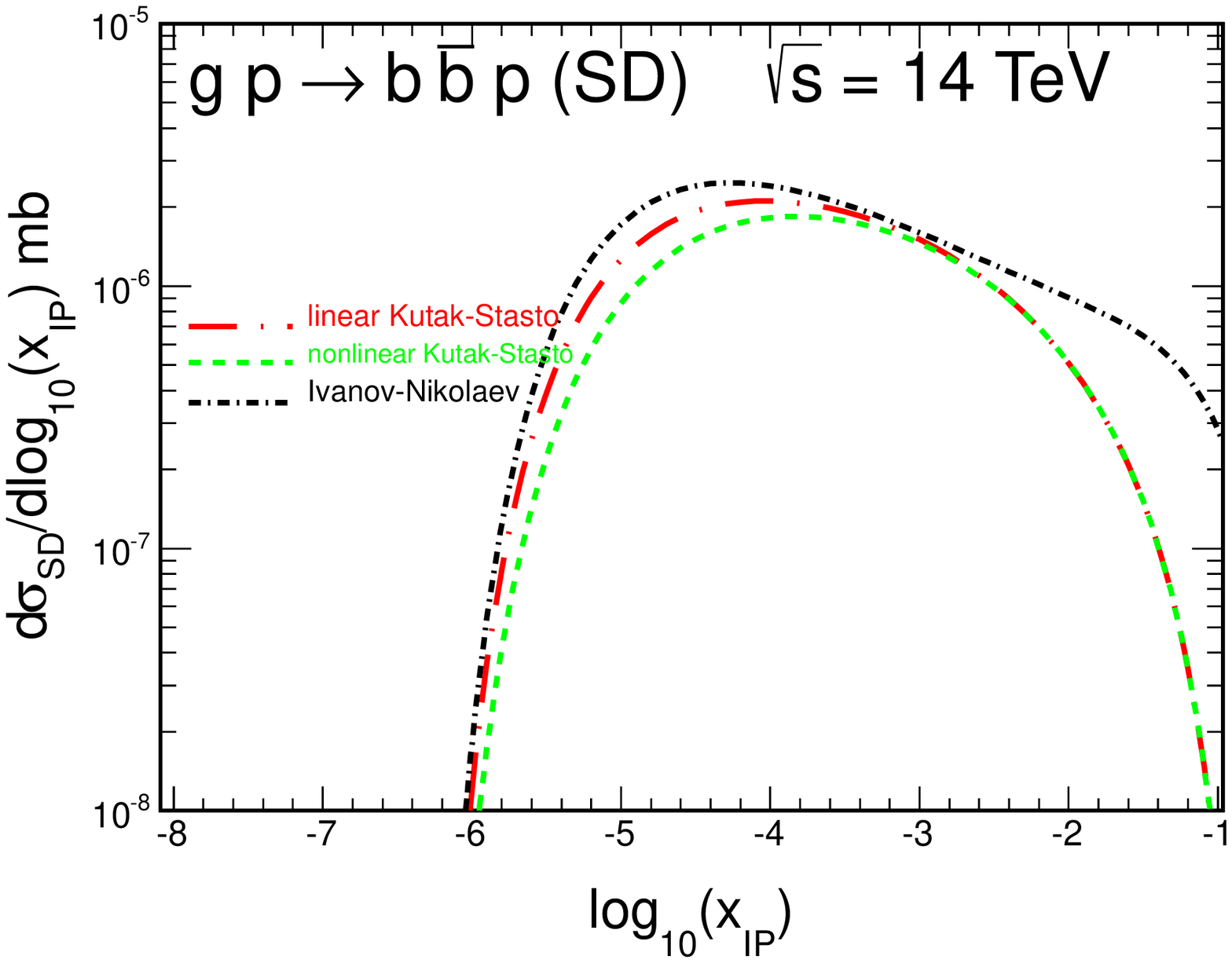}
   \caption{
\small Distribution in log$_{10}(x_{\Pom})$ for $c \bar c$ (left) and 
$b \bar b$ (right) produced in a single diffractive process for 
center of mass energy $\sqrt{s}$ = 14 TeV for the Ivanov-Nikolaev (solid),
linear Kutak-Sta\'sto (dashed) and nonlinear Kutak-Sta\'sto (dotted) UGDFs. 
Absorptive effects have been included by multiplying by
gap survival factor.
}
\label{fig:dsig_dxi2}
\end{figure}

Now we wish to present differential distributions.
Here, absorption corrections 
are included in a rough manner, by multiplying the cross section by a gap survival factor $S_G$ = 0.05
\cite{KMR_eikonal,Maor}. 
A more subtle treatment, which would include the dependence of absorption effects 
on kinematical variables goes beyond the scope of the present work, but
must be developed in the future.

Let us start with distributions in $x_{\Pom}$ -- the fractional longitudinal
momentum loss of proton. Notice that $\log(1/x_\Pom)$ is proportional to the size of 
the rapidity gap. The cross section drops sharply at $x_\Pom \lsim 10^{-7}$ for charm
quarks and $x_\Pom \lsim 10^{-6}$ for bottom quarks. This is related to the fact
that with increasing gap size we are asking for harder partons in the dissociating
proton. The gluon distribution however drops sharply at large $x$.

Notice, that in the Ingelman-Schlein model, the gap size-dependence
is described in terms of a universal flux of Pomerons.  
In our microscopic model the $x_\Pom$--dependence is driven by the 
dependence of the unintegrated gluon distribution on $x_{\mathrm{eff}}=x_\Pom$. 

We observe a breaking of Regge factorization: the gap-size dependence is 
substantially steeper for bottom production than for charm production. It reminds us 
of the systematics of the energy dependence of diffractive photo/leptoproduction of
vector mesons as well as inclusive deep inelastic diffraction, 
where also a dependence of the effective Pomeron intercept on the 
relevant hard scale is observed.
\begin{figure}[!h]
\includegraphics[width=7.5cm]{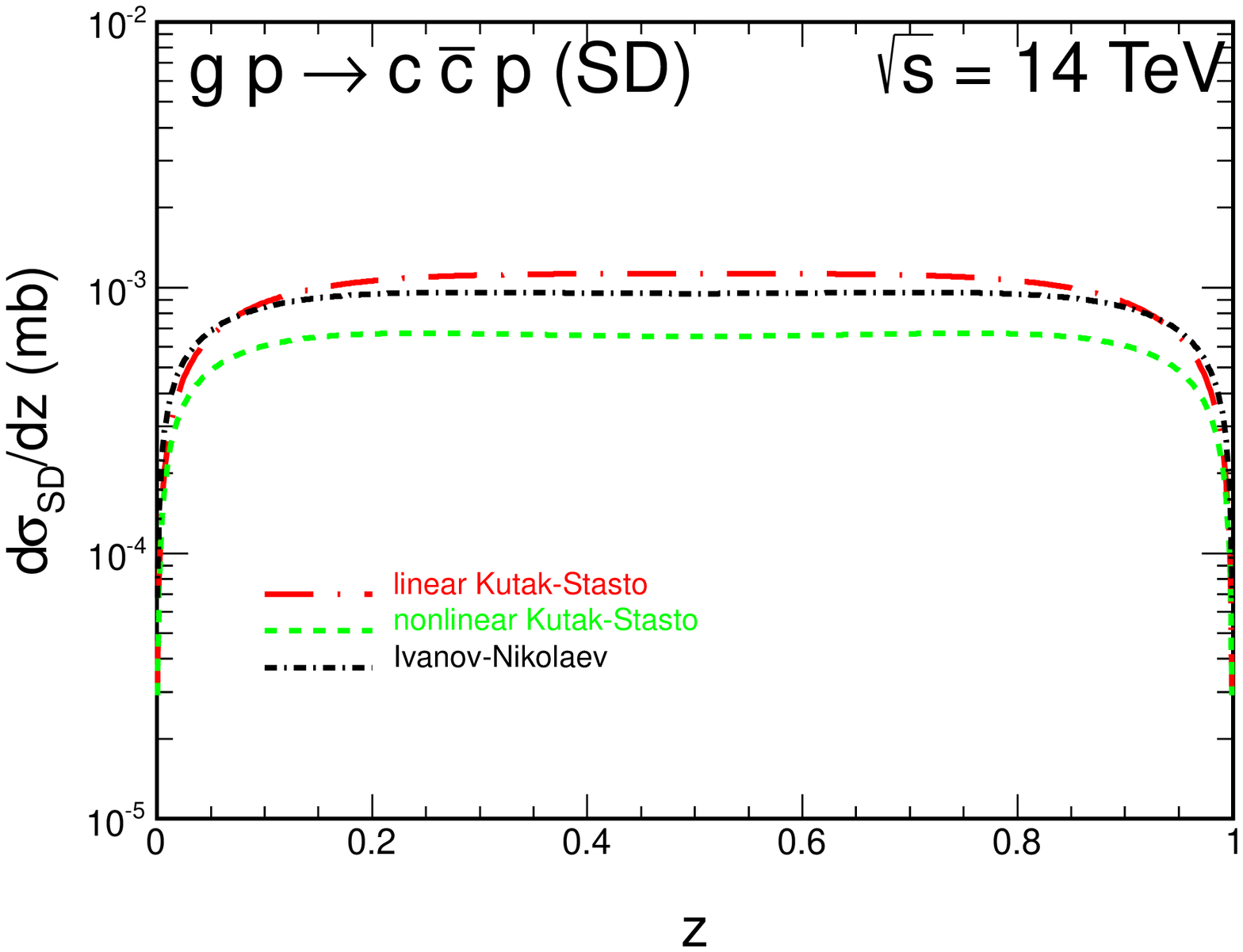}
\includegraphics[width=7.5cm]{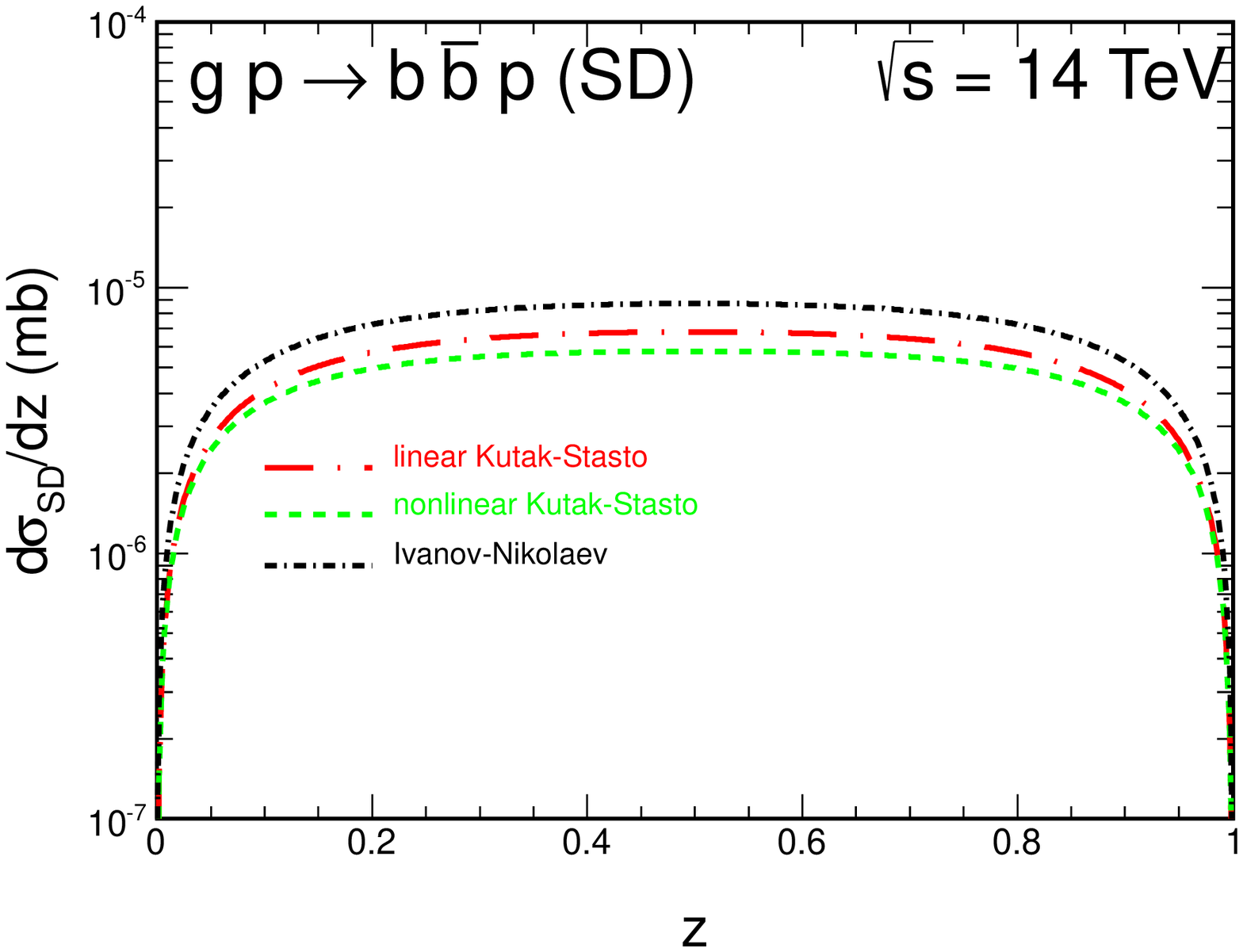}
   \caption{
\small Distribution in $z$ of $c$ ($\bar c$) (left) and $b$ ($\bar b$) 
(right) produced in a single diffractive process for center of mass 
energy $\sqrt{s}$ = 14 TeV for the Ivanov-Nikolaev and the Kutak-Sta\'sto 
UGDFs. Absorptive effects have been included by multiplying the model
cross section by gap survival factor.
}
\label{fig:dsig_dz}
\end{figure}
In Fig.\ref{fig:dsig_dz} we show a distribution in longitudinal momentum
fraction of quark with respect to the parent gluon. It is fairly uneventful
and simply reflects the $z$--dependence in evidence in Eq.(\ref{eq:parton-level}).
The distribution is broad, it vanishes at the endpoints, because there the invariant 
mass of the $Q \bar Q$ system becomes large, and the unintegrated gluon distribution
drops steeply at $x_\mathrm{eff} \to 1$.

\begin{figure}[!h]
\includegraphics[width=7.5cm]{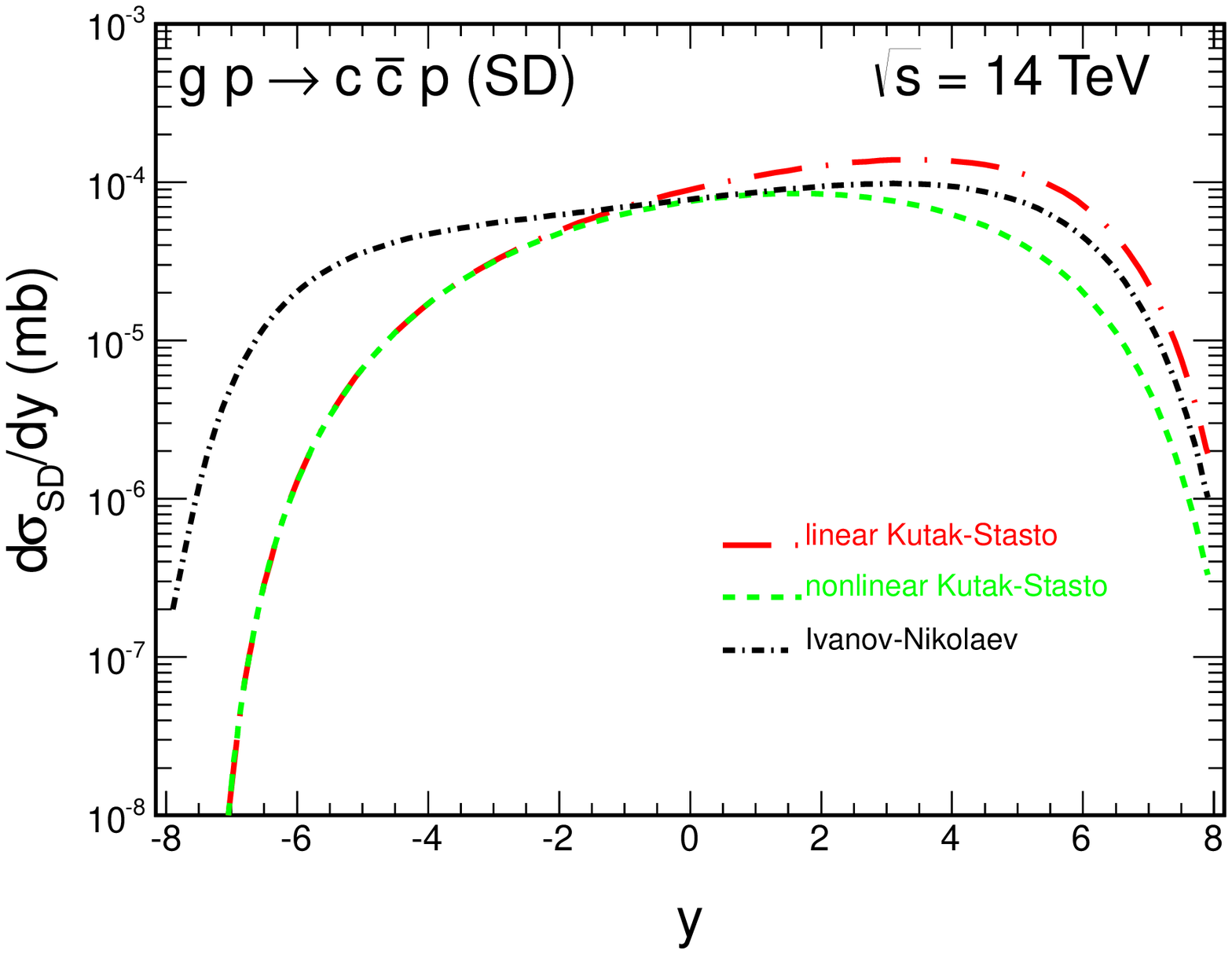}
\includegraphics[width=7.5cm]{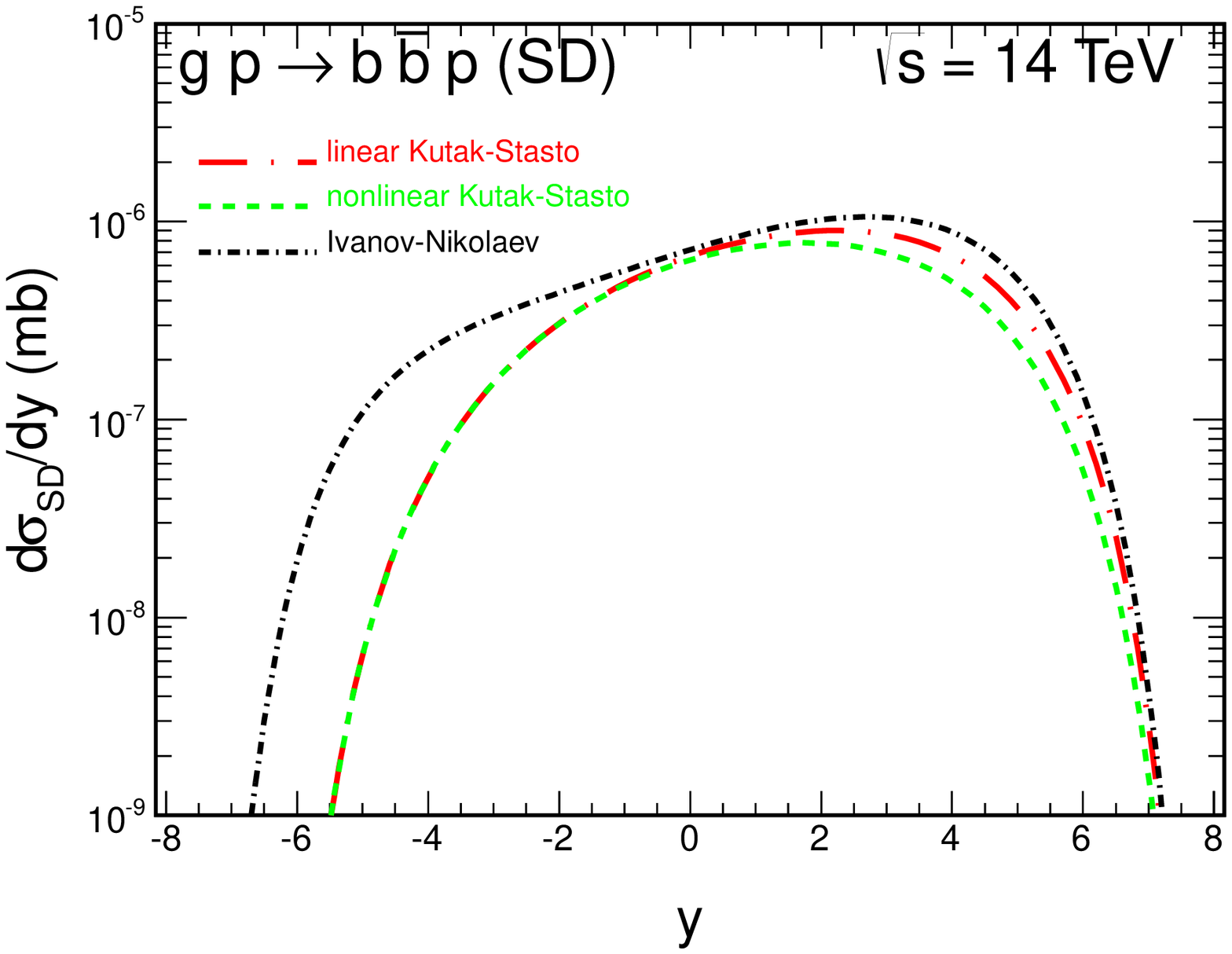}
   \caption{
\small Distribution in rapidity of $c$ ($\bar c$) (left) and $b$ ($\bar
b$) (right) produced in a single diffractive process for center of mass 
energy $\sqrt{s}$ = 14 TeV.
Absorptive effects have been included by multiplying by
gap survival factor.
}
\label{fig:dsig_dy}
\end{figure}

In Fig.\ref{fig:dsig_dy} we present the rapidity distribution of charm (left
panel) and bottom (right panel) quarks/antiquarks from diagram (b)
in Fig.\ref{fig:sd_diagrams}. 
At large rapidities the cross section with the Ivanov-Nikolaev UGDF
is much larger than that with the nonlinear Kutak-Sta\'sto UGDF. 
This is partially due to nonlinear effects included in the latter 
distributions. The nonlinear effects show up at $x_\Pom <$ 10$^{-4}$.

\begin{figure}[!h]
\includegraphics[width=7.5cm]{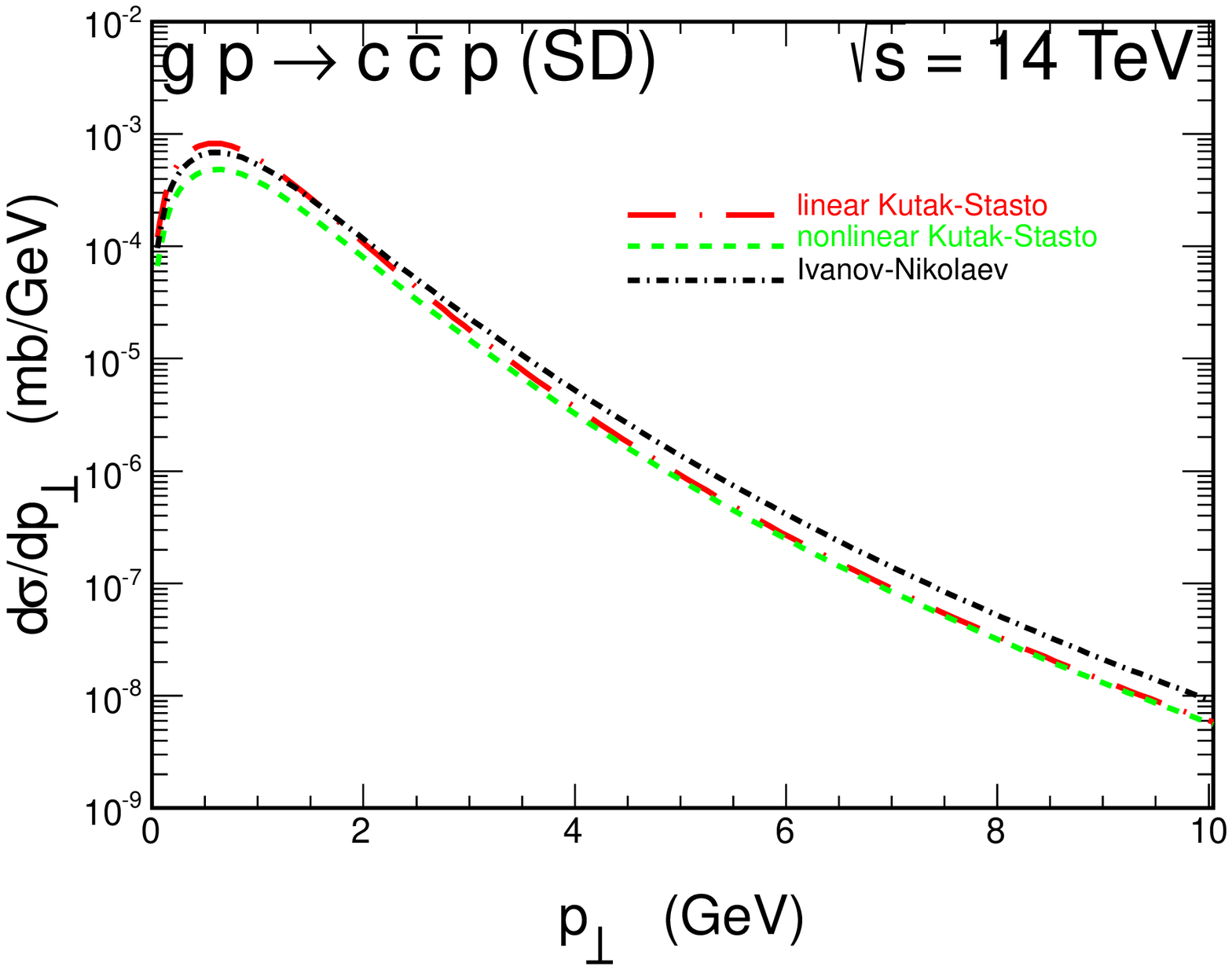}
\includegraphics[width=7.5cm]{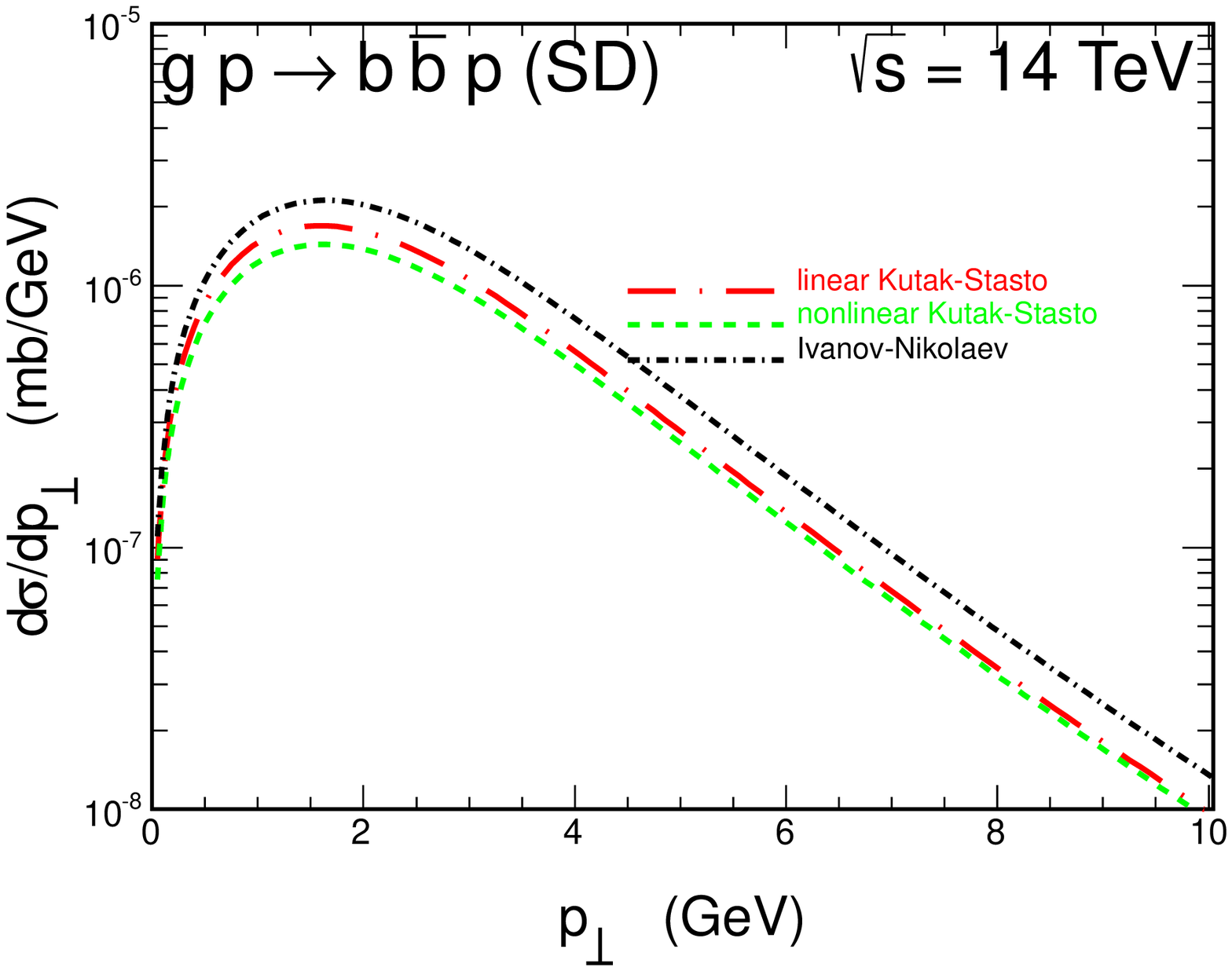}
   \caption{
\small Distribution in transverse momentum of $c$ ($\bar c$) (left) 
and $b$ ($\bar b$) (right) produced in a single diffractive process 
for center-of-mass energy $\sqrt{s}$ = 14 TeV.
Absorptive effects have been included by multiplying by
gap survival factor.
}
\label{fig:dsig_dpt}
\end{figure}

\begin{figure}[!h]
\includegraphics[width=7.5cm]{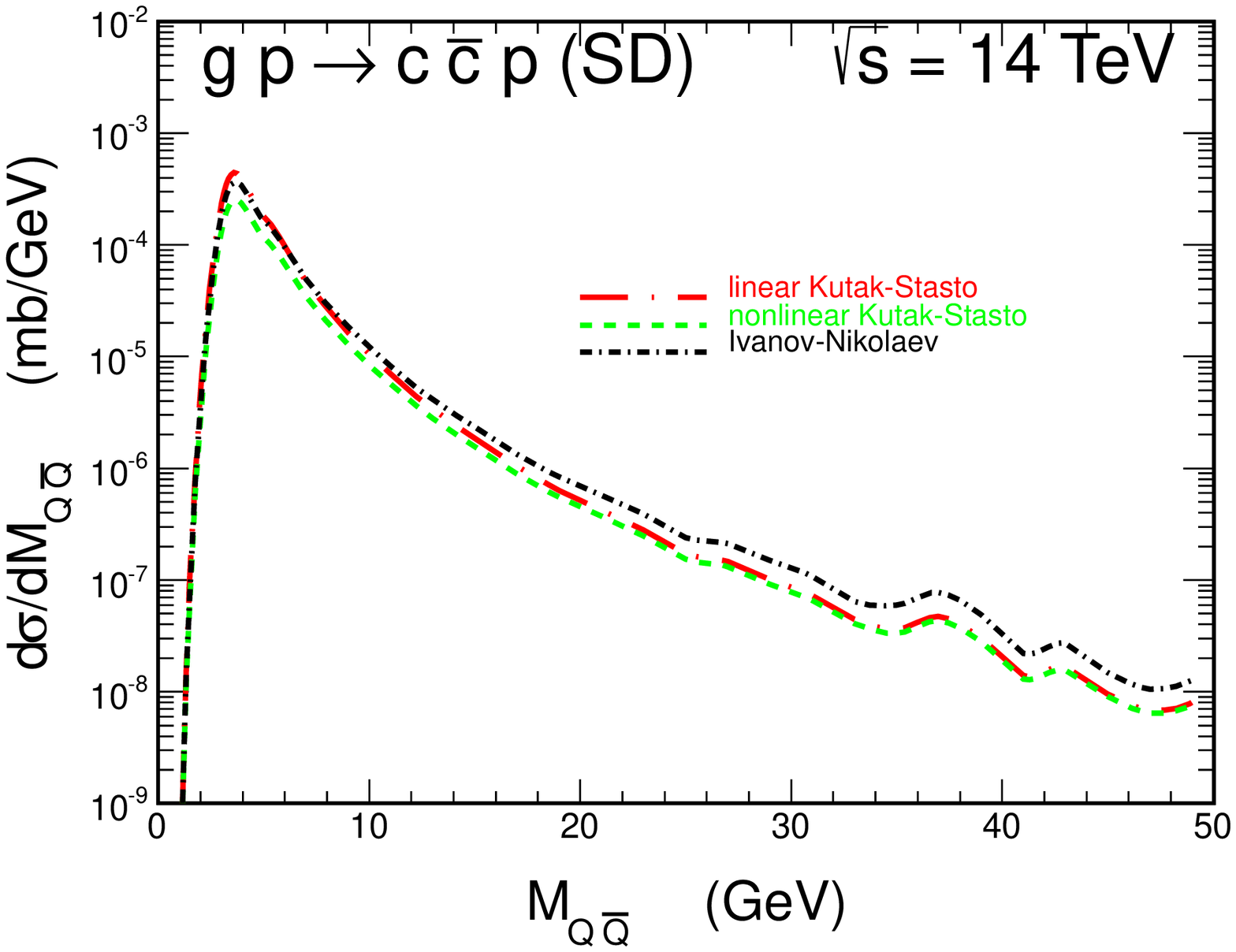}
\includegraphics[width=7.5cm]{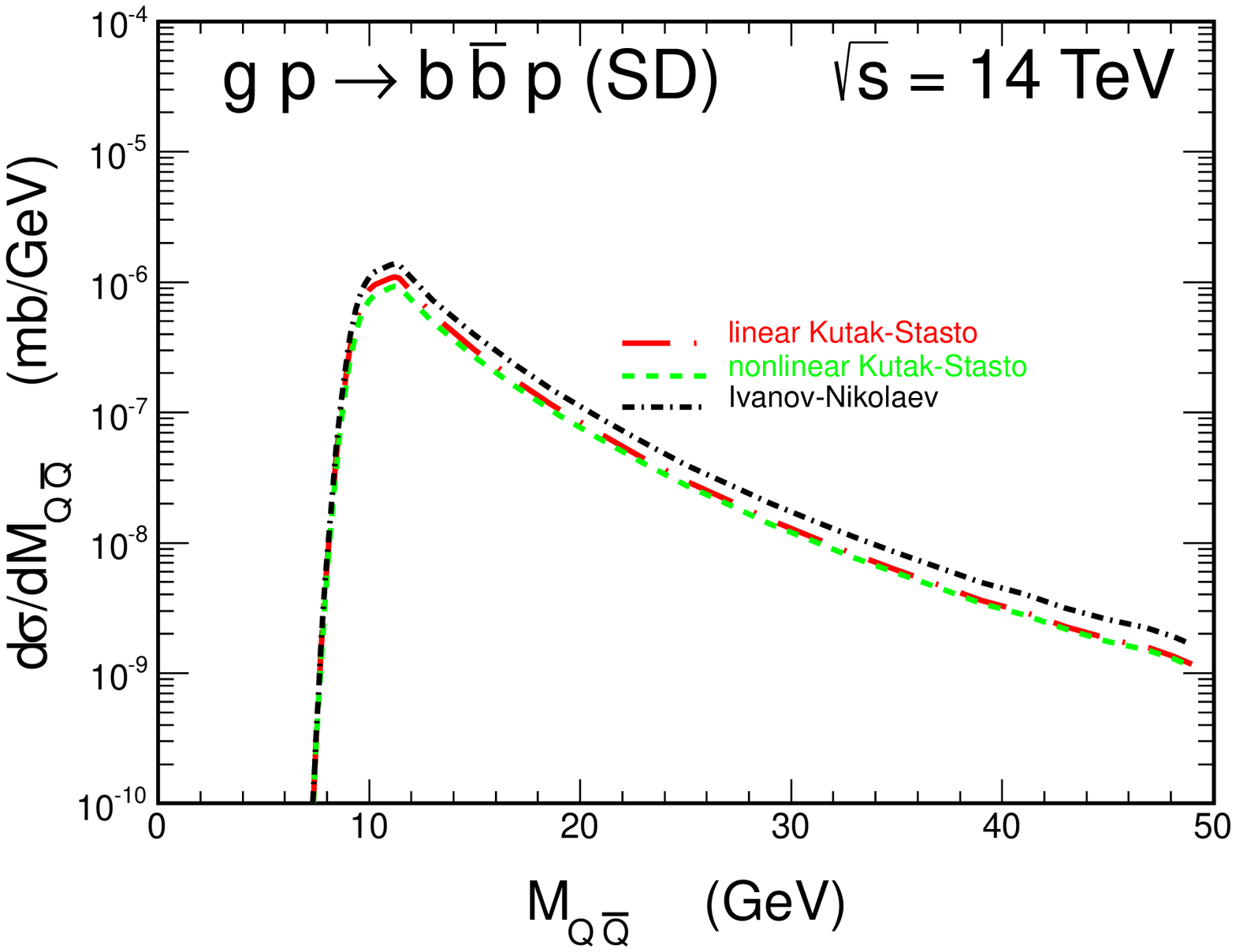}
   \caption{
\small Distribution in invariant mass of $c \bar c$ (left) 
and $b \bar b$ (right) system produced in a single diffractive process 
for center of mass energy $\sqrt{s}$ = 14 TeV.
Absorptive effects have been included by multiplying by
gap survival factor.
}
\label{fig:dsig_dMqqbar}
\end{figure}

In Fig.\ref{fig:dsig_dpt} we show transverse momentum distributions of 
charm (left panel) and bottom (right panel) quarks/antiquarks from
one single-diffractive mechanism. The spread in transverse momentum here
is somewhat smaller than in the Ingelman-Schlein model calculations of Ref.\cite{LMS2011}.
Finally in Fig.\ref{fig:dsig_dMqqbar} we show distribution in the 
quark-antiquark invariant mass. Rather low invariant masses (small
rapidity difference between quark and antiquark) close to the $Q \bar Q$
threshold, especially for $c \bar c$ production, are populated.

\begin{figure}[!h]
\includegraphics[width=9cm]{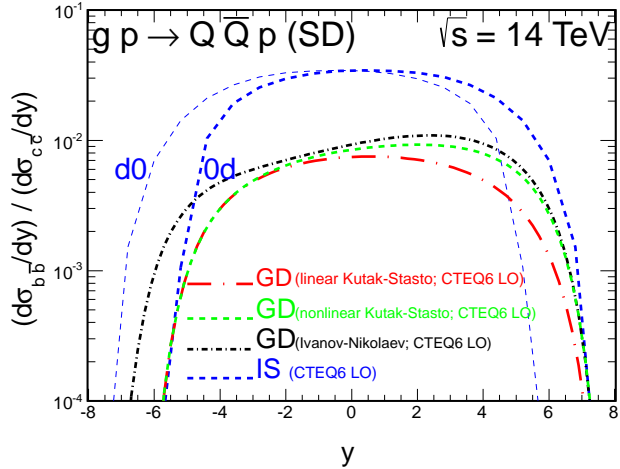}
   \caption{
\small The ratio of the $b \bar b$ to $c \bar c$ distributions in 
quark (antiquark) rapidity.
}
\label{fig:dsig_dy_ratio}
\end{figure}

\begin{figure}[!h]
\includegraphics[width=9cm]{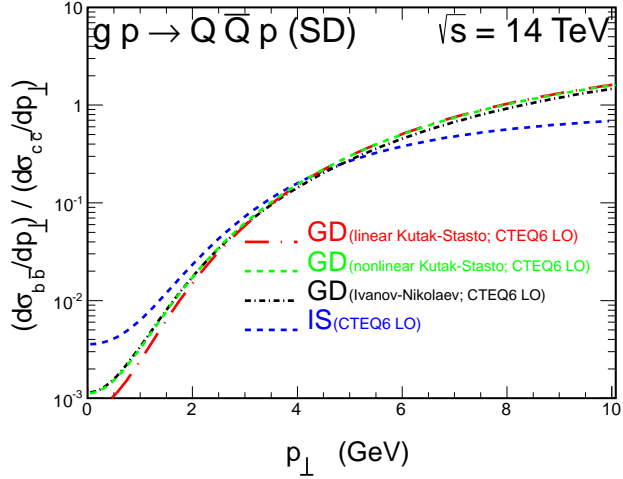}
   \caption{
\small The ratio of the $b \bar b$ to $c \bar c$ distributions in 
quark (antiquark) transverse momentum.
}
\label{fig:dsig_dpt_ratio}
\end{figure}

\begin{figure}[!h]
\includegraphics[width=9cm]{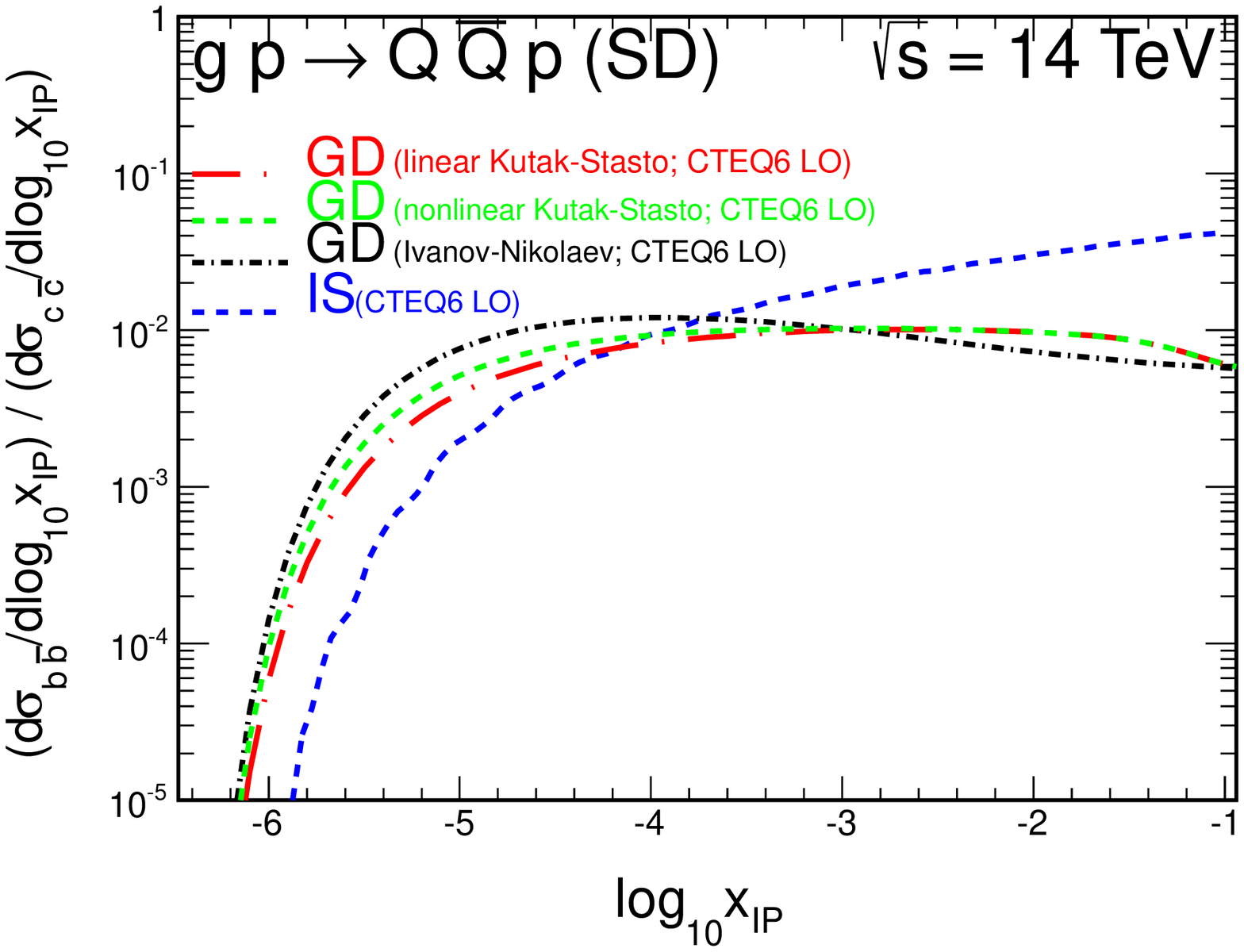}
   \caption{
\small The ratio of the $b \bar b$ to $c \bar c$ distributions 
in $\log_{10}(x_{\Pom})$.
}
\label{fig:dsig_dlog10xpom_ratio}
\end{figure}

\begin{figure}[!h]
\includegraphics[width=9cm]{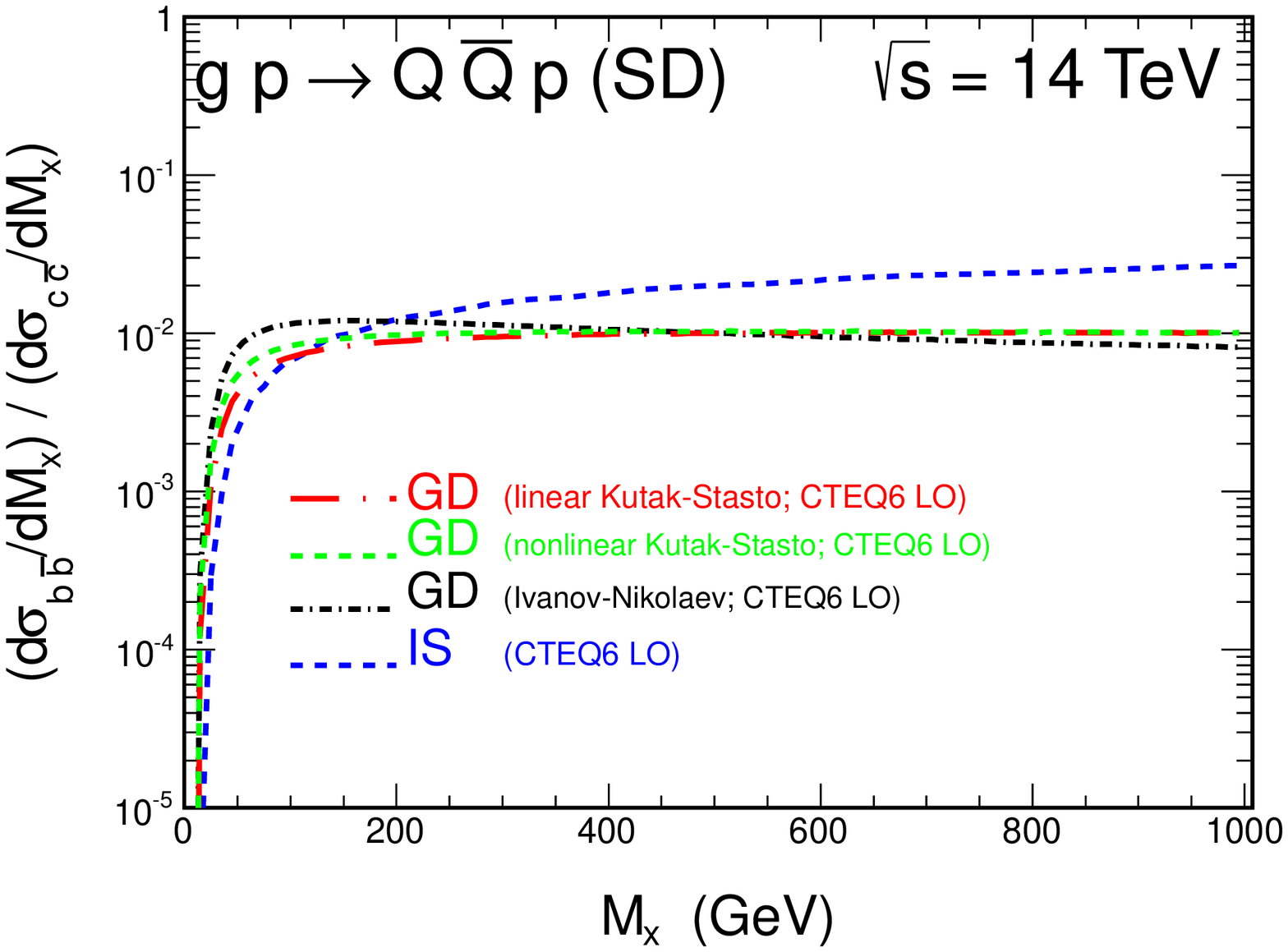}
   \caption{
\small The ratio of the $b \bar b$ to $c \bar c$ distributions 
in $M_X$.
}
\label{fig:dsig_dMX_ratio}
\end{figure}

Now we wish to study dependence of the ratio of cross sections
for $b \bar b$ and $c \bar c$ production as a function of several
kinematical variables, which should be to a good approximation
independent of absorption. In Fig.\ref{fig:dsig_dy_ratio} we show
the ratio as a function of quark rapidity. The ratio for the
Ingelman-Schlein model is somewhat larger than that for the
gluon-dissociation approach.
The charm-to-bottom ratio as a function of transverse momentum of the (anti)quark is shown in 
Fig.\ref{fig:dsig_dpt_ratio}. The ratio increases as a function of
quark transverse momentum. The character of the function is in principle
similar for both models.
In Fig.\ref{fig:dsig_dlog10xpom_ratio} we show similar ratio as a
function of log$_{10}x_{\Pom}$. Here we see that the gluon dissociation
mechanism and the Ingelman-Schlein model exhibit an opposite trend with inreasing 
gap size: the $b \bar b$ fraction increases for the gluon dissociation model and
decreases for the Ingelman-Schlein case.
Finally, in Fig.\ref{fig:dsig_dMX_ratio} we show the ratio as a function of diffractively produced mass
$M_X$. The dependences for both models are rather smooth. 
\section{Conclusions and outlook}
In the present paper we have discussed forward amplitudes for the 
$g p \to Q \bar Q p$ subprocess both in the impact parameter and
momentum space representation in the forward scattering approximation.
The amplitude for the off-forward directions within the diffraction cone
was extrapolated by assuming exponential dependence known from other diffractive processes.
The forward amplitude for the $g p \to Q \bar Q p$ subprocess has been obtained 
in terms of unintegrated gluon distribution of the target proton. 

The formulae have been used to calculate cross section for the single 
scattering process $p p \to Q \bar Q p X$ as a convolution of the
collinear gluon distributions in the proton and the elementary 
$g p \to Q \bar Q p$ cross section both for charm and bottom production.
When applied to the hadronic collisions, this approach allows one to predict 
heavy quark production ``close to the gap''. In other words the heavy quarks
are produced in the Pomeron fragmentation region.

Three different unintegrated gluon distributions describing deep-inelastic data
at HERA, known from the literature, have been used.

We have presented first results for the rapidity and transverse momentum
distribution of quarks (antiquarks) and $Q \bar Q$ invariant mass
at the nominal LHC energy $\sqrt{s}$ = 14 TeV. The cross section
for charm quarks is two orders of magnitude larger than that for bottom 
quarks, as expected from the $m_Q^{-4}$ scaling of the partonic subprocess.

We have calculated also ratio of the cross section for $b \bar b$
and $c \bar c$ as a function of several kinematical variables.
The ratio is fairly smooth in (anti)quark rapidity and strongly depends
on (anti)quark transverse momentum. 

It would be interesting to extend the present microscopic model
to higher Fock-states, like $Q \bar Q g$. This would make it possible
to also discuss heavy quark production ``far away from the gap'', which
up to now is modelled in the Ingelman-Schlein approach.
We believe that a more microscopic approach may make it possible to
find sources of Regge factorization breaking, and perhaps to ultimately
include absorption effects on an equal footing.

A measurement of the single diffractive production would be possible
e.g. at ATLAS detector by using so-called ALFA detectors for measuring 
forward protons and their fractional energy loss and the main central 
detector for the measurement of $D$ or $B$ mesons. CMS+TOTEM is another 
option.
Further evaluation of the cross section including $c \to D$ or $b \to B$
fragmentation and experimental cuts on pseudorapidities and transverse
momenta of $D$ mesons and protons) would be very useful in planning 
and performing measurements. Such measurements should be possible after
the technical shut down in 2013-2014.

\vspace{1cm}

{\bf Acknowledgment}

This work was partially supported by the Polish MNiSW grant 
DEC-2011/01/B/ST2/04535.



\end{document}